\RequirePackage[T1]{fontenc}
\documentclass[12pt]{article}
\pdfoutput=1 % if your are submitting a pdflatex (i.e. if you have images in pdf, png or jpg format)
\usepackage[height=8.85in,width=6.45in]{geometry}

\usepackage[utf8]{inputenc}
\usepackage{amsmath}
\usepackage{amssymb}
\usepackage{mathtools}
\numberwithin{equation}{section}
\usepackage{slashed}
\usepackage{braket}
\usepackage[svgnames]{xcolor}
\usepackage[colorlinks,linktocpage=true,citecolor=DarkGreen,linkcolor=FireBrick]{hyperref}
\usepackage{cite}
\usepackage{graphicx}
\usepackage{tikz}
\usepackage{tikz-cd}
\usepackage{times}
\usepackage[scaled]{couriers}
\usepackage{bm}
\usepackage{subfig}
\usepackage{mathrsfs}
\usepackage{amsthm}

\usepackage{xcolor}
\usepackage{mdframed}

\usepackage{slashed}

\def\beq#1\eeq{\begin{align}#1\end{align}}

\newcommand{\GeV}{\ {\rm GeV} }

\newcommand{\lmk}{\left(} 
\newcommand{\rmk}{\right)}
\newcommand{\lkk}{\left[} 
\newcommand{\rkk}{\right]}

\newcommand{\del}{\partial}

\newcommand{\Mpl}{M_{\rm Pl}}
\newcommand{\abs}[1]{\left\vert {#1} \right\vert}

\newcommand{\eq}[1]{Eq.~(\ref{#1})}
\newcommand{\eqs}[1]{Eqs.~(\ref{#1})}

\begin{document}

\begin{titlepage}

\begin{flushright}
TU-1276
\end{flushright}

\vskip 3cm

\begin{center}

{\Large \bfseries Analytic derivation of GW spectrum\\ \vspace{0.4cm}
from bubble collisions in FLRW Universe}

\vskip 1cm

Masaki~Yamada

\vskip 1cm

\begin{tabular}{ll}
Department of Physics, Tohoku University, Sendai 980-8578, Japan
\end{tabular}

\vskip 1cm

\end{center}

\noindent
We generalize the analytic formula for the gravitational-wave spectrum from bubble collisions during a cosmological first-order phase transition, under the thin-wall and envelope approximations, by incorporating the effect of cosmic expansion in the FLRW metric. Along with presenting the complete analytic expression and corresponding numerical results, we also derive simplified formulas valid in the large- and small-$k$ limits, as well as in the Minkovski limit. The latter expansion reveals that the Minkovski approximation breaks down for $\beta / H_* \lesssim 10$, where $\beta$ denotes the inverse duration of the phase transition and $H_*$ the Hubble parameter at its completion. Furthermore, the next-to-leading-order term contributes about a $10\%$ correction for $\beta / H_* \sim 140$, a typical value for the electroweak phase transition.

\end{titlepage}

\setcounter{tocdepth}{2}

\newpage

\tableofcontents

\section{Introduction}

Cosmological first-order phase transitions (FOPTs) are compelling targets for gravitational-wave (GW) astronomy. As the Universe expands and cools, a hot plasma may undergo a phase transition driven by physics beyond the Standard Model, often associated with the dynamics of a scalar field and spontaneous symmetry breaking~\cite{Quiros:1999jp,Morrissey:2012db,Schwaller:2015tja,Athron:2023xlk}. FOPTs proceed via the nucleation of bubbles of the true vacuum that expand, collide, and stir the surrounding plasma. Because the relevant length scales can be macroscopic, the associated out-of-equilibrium dynamics can source a stochastic GW background potentially observable with current and planned detectors~\cite{Janssen:2014dka,LISA:2017pwj,Kawamura:2011zz,Kawamura:2020pcg,Harry:2006fi,Punturo:2010zz,Maggiore:2019uih,Reitze:2019iox,Somiya:2011np,KAGRA:2020cvd,Colpi:2024xhw}. 
The resulting GW signal carries information about the underlying particle-physics scales and couplings, offering a unique window into sectors that may be otherwise inaccessible to terrestrial experiments.

Three mechanisms mainly contribute to the GW spectrum produced by a FOPT: (i) collisions of scalar-field bubble walls (often modeled with the thin-wall/envelope approximation), (ii) long-lived acoustic (sound-wave) modes in the plasma, and (iii) magnetohydrodynamic turbulence. When friction between bubble walls and the ambient plasma is significant, most of the released vacuum energy goes into bulk fluid motion, and the sound-wave contribution typically dominates~\cite{Hindmarsh:2013xza,Hindmarsh:2017gnf,Hindmarsh:2019phv,Caprini:2009yp,RoperPol:2019wvy}. Quantitative predictions in this regime rely on large-scale numerical simulations, which are computationally costly and make it difficult to map spectral features across wide regions of parameter space. In contrast, when the interaction with the plasma is sufficiently weak, the bubble walls “run away,” accelerating toward the speed of light. In this regime the GW signal is dominated by the scalar-field (bubble-wall) contribution, which admits controlled semi-analytic treatments and substantially reduces numerical cost~\cite{Kosowsky:1992rz,Kosowsky:1992vn,Kamionkowski:1993fg,Huber:2008hg,Bodeker:2009qy,Weir:2016tov,Jinno:2016vai,Konstandin:2017sat,Jinno:2017fby,Cutting:2018tjt,Megevand:2021juo,Cai:2023guc}.

Recent interest has focused on supercooled FOPTs, which arise naturally in theories with approximate conformal symmetry (e.g., a dilaton-like or dark-sector scalar)\cite{Ellis:2019oqb,Lewicki:2019gmv,Lewicki:2020jiv,Ellis:2020nnr,Lewicki:2022pdb} (see also\cite{Konstandin:2011dr,Megevand:2016lpr,Ellis:2018mja,Hashino:2018wee,Brdar:2018num,Fujikura:2019oyi}). Such transitions may occur in a hidden sector that is decoupled from the visible plasma, making the runaway-bubble scenario particularly well motivated. An important complication for supercooled transitions is that their characteristic duration can be comparable to the Hubble time, so that cosmic expansion can noticeably modify both the amplitude and the spectral shape of the resulting GW background. In this situation, formulae in Minkovski spacetime become inadequate, and a treatment that consistently incorporates the expanding Friedmann–Lemaître–Robertson–Walker (FLRW) background is required.

In this work, we extend the analytic bubble-collision GW formula of Ref.\cite{Jinno:2016vai} from Minkovski spacetime to an expanding FLRW Universe, under the thin-wall and envelope approximation.\footnote{Related recent work investigates gravitational effects on bubble-driven fluid profiles and acoustic GW production when the mean bubble spacing is a non-negligible fraction of the Hubble radius~\cite{Jinno:2024nwb,Giombi:2025tkv}. These effects become especially relevant for finite-width sound shells.} Throughout, we neglect the backreaction of the transition on the background expansion (i.e., the bubble energy density remains subdominant), but we fully include (i) the time-dependent nucleation rate, (ii) a possible time-dependent vacuum energy, and (iii) the effects of the scale factor on bubble growth and GW propagation.\footnote{A closely related study\cite{Zhong:2021hgo} also extends analytic formulas to an FLRW background. They find that Minkovski assumptions can overestimate GW amplitudes for small $\beta/H$. Their analysis, however, implicitly adopts specific time dependences for the vacuum energy and nucleation rate. Moreover, their definition of $\alpha$ (vacuum-to-total energy ratio) is not tied to an explicit time slice, which becomes crucial when the transition duration approaches the Hubble time.} Our formulation is naturally expressed in conformal time and comoving coordinates; the resulting expressions closely parallel those in Minkovski spacetime, with physical lengths, times, and parameters replaced by their conformal counterparts. Importantly, we do not assume radiation domination: the formula applies to an arbitrary expansion history and to a generic nucleation history.

The extended formula makes several consequences transparent. We derive controlled asymptotics in the small- and large-wavenumber limits, which clarify how the spectral slopes are deformed by expansion.%
\footnote{
See also Ref.~\cite{Megevand:2021llq} for analytical study of the asymptotic behavior in a Minkovski background.
}
The resulting integrals are numerically well behaved, so multi-dimensional integrations can be performed efficiently. In addition, we develop a systematic expansion around the Minkovski limit in powers of the small parameter $\epsilon \equiv H/\beta$, quantifying the leading corrections due to cosmic expansion. We also compute the full spectrum numerically and compare it with the asymptotic expressions, delineating the domain of validity of the Minkovski approximation and identifying when next-to-leading-order terms in $\epsilon$ become non-negligible. These results provide a fast and flexible way to explore the GW phenomenology of runaway, supercooled transitions across broad classes of hidden-sector models and cosmological backgrounds.

This paper is organized as follows.
In Sec.\ref{sec:expansion}, we clarify the definitions of the physical quantities and parameters used throughout this work, emphasizing which of them can be time dependent in an expanding background.
In Sec.\ref{sec:analytic}, we derive analytic expressions for the GW spectrum. There are two distinct contributions, called single-bubble and double-bubble contributions. While the latter is typically subdominant, we derive formulas for both.
In Sec.\ref{sec:asymptotic}, we analyze the asymptotic regimes at large/small wavenumbers and simplify the analytic formula accordingly.
In Sec.\ref{sec:largebeta}, we perform an expansion around the Minkowski limit, which clarifies when the calculation in Minkovski spacetime is justified and when its approximation is violated.
In Sec.\ref{sec:numerical}, we present numerical results for several scenarios, focusing especially on the radiation-dominated epoch, and compare them with the asymptotic formulas.
In Sec.\ref{sec:implication}, we show how the peak amplitude depends on the nucleation history and demonstrate that the GW amplitude admits an upper bound even for supercooled transitions.
Finally, Sec.~\ref{sec:discussion} concludes.

\section{Assumptions and approximations}
\label{sec:expansion}

We consider GW emission from a FOPT in a FLRW Universe. The spacetime metric is given by
\begin{align}
ds^2
&= a^2(\tau) \lkk - d\tau^2 + (\delta_{ij}+2h_{ij})dx^idx^j \rkk \,,
\end{align}
where $a(\tau)$ is the scale factor, $\tau$ is the conformal time, and $h_{ij}$ denotes the tensor perturbations. The tensor perturbations satisfy the transverse-traceless conditions $\del_i h_{ij} = h_{ii} = 0$. We treat $h_{ij}$ as a small perturbation around the FLRW background and compute its spectrum generated by vacuum-bubble collisions. Throughout this paper, we adopt the Fourier transformation conventions $\int d^3x \; e^{i\vec{k}\cdot\vec{x}}$ and $\int d^3k/(2\pi)^3 \; e^{-i\vec{k}\cdot\vec{x}}$, where $k$ denotes the comoving momentum.

We consider a scenario in which the Universe is initially trapped in a false vacuum at high temperature and undergoes a FOPT as the temperature falls below a critical value. This transition proceeds via the nucleation of bubbles of true vacuum, which subsequently expand and collide until the entire Universe transitions to the true vacuum state. Throughout this work, we assume that the velocity of the bubble walls is as high as the speed of light.

We define the bubble nucleation rate per unit {\it comoving} volume per unit {\it conformal} time as 
\beq
 \tilde{\Gamma}(\tau) \equiv a^4 (\tau) \Gamma(\tau) \,,
 \label{eq:gammatilde}
\eeq
where $\Gamma$ is the nucleation rate per unit physical volume per unit physical time.

The probability that a given spatial point remains in the false vacuum at conformal time $\tau$ is given by $P_1(\tau) = e^{-I_1(\tau)}$, where 
\beq
 &I_1(\tau) 
 = \int_0^{\tau} d\tau' 
 \frac{4\pi}{3} \lmk \tau - \tau' \rmk^3 \tilde{\Gamma} (\tau') \,.
 \label{eq:I}
\eeq
We define the time of bubble collision, $\tau_*$, as the moment when the false-vacuum survival probability becomes $\mathcal{O}(1)$, i.e., when $I_1(\tau_*) = 1$. The Hubble parameter and conformal Hubble parameter evaluated at $\tau = \tau_*$ are denoted by $H_*$ and $\mathcal{H}_*$ ($= a(\tau_*) H_*$), respectively.

We quantify the (conformal) growth rate of the true vacuum fraction at $\tau = \tau_*$ by 
\begin{align}
\tilde{\beta} \equiv - \frac{d \ln P_1}{d\tau}(\tau_*) = \frac{d I_1}{d \tau} (\tau_*)\,.
\label{eq:beta}
\end{align}
We also define $\beta \equiv \tilde{\beta}/a(\tau_*)$, which corresponds to the inverse of the (physical) duration of the phase transition.
In particular, the ratio $\tilde{\beta} / \mathcal{H}_*$ (equivalently, $\beta / H_*$) characterizes the rapidity of the phase transition relative to the Hubble expansion rate.
In the limit $\tilde{\beta} / \mathcal{H}_* \gg 1$, one expects the effect of Hubble expansion to be negligible.

We denote the difference in energy density between the false vacuum and the true vacuum (or, more precisely, the potential energy at the tunneling point) by $\rho_0(\tau)$.
We consider the general case where $\rho_0$ exhibits time dependence.

We define a parameter $\alpha_*$ as the ratio of the energy density of the false vacuum to that of background components in the Universe $\rho_{\rm tot}(\tau)$ ($=3 H^2(\tau) \Mpl^2$) at $\tau = \tau_*$: 
\begin{align}
\alpha_*
&\equiv \frac{\rho_0 (\tau_*)}{\rho_{\rm tot}(\tau_*)} \,.
\label{eq:alpha}
\end{align}
Note that this ratio is generally time-dependent; we therefore specify its value at $\tau = \tau_*$ by introducing the notation $\alpha_*$.
This parameter characterizes the typical magnitude of the energy density released during the phase transition.% 
\footnote{
\label{footnote1}
Note that $\alpha_*$ is not identical to the latent heat. It may be defined by 
\beq
 \alpha_{\rm b} = \int d \tau \frac{dP_1(\tau)}{d \tau} 
 \frac{\rho_0 (\tau)}{\rho_{\rm b,rad}(\tau)} \,,
\eeq
with $\rho_{\rm b,rad}$ being the energy density of radiation in the bubble-nucleation sector. 
This quantity can be much larger than unity and hence we can consider runnaway bubbles even for the case with $\alpha_* \ll 1$, assuming $\rho_{\rm tot} \gg \rho_{\rm b, rad}$. 
}

Throughout this paper, we neglect the backreaction of bubble nucleation on the metric and approximate the background spacetime as a homogeneous and isotropic FLRW Universe. This approximation is justified when $\alpha_* \ll 1$.% 
\footnote{If $\alpha_*$ is of order $0.01$ or larger, some regions remain in the false vacuum and undergo eternal inflation in the case of $\tilde{\beta}/\mathcal{H}_* \simeq 3$. From the perspective of an external observer, these regions appear as black holes. In Ref.~\cite{Jinno:2023vnr}, we discussed that this mechanism leads to overproduction of black holes when $\alpha_* \gtrsim 0.01$.}

\begin{figure}
 \centering
 \includegraphics[width=0.45\hsize]{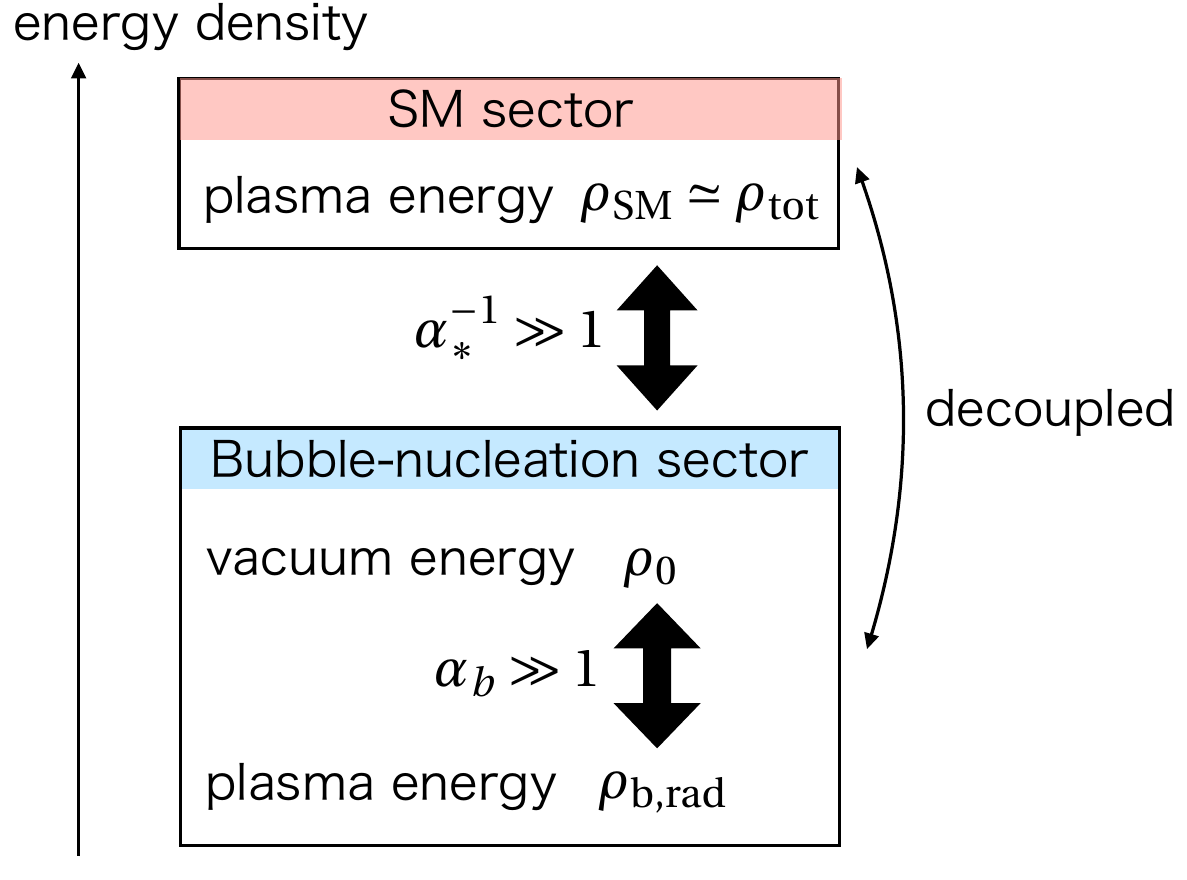}
 \caption{
Schematic illustration of the hierarchy among energy densities.
The SM energy density $\rho_{\mathrm{SM}}$ ($\simeq \rho_{\rm tot}$) dominates the total energy of the Universe and determines the Hubble expansion rate.
The vacuum energy in the bubble-nucleation sector $\rho_0$ is much smaller than $\rho_{\mathrm{SM}}$, ensuring $\alpha_* \ll 1$. However, within the bubble-nucleation sector, $\rho_0$ dominates over its plasma energy $\rho_{\mathrm{b, rad}}$, allowing the phase transition to be strongly supercooled.
}
 \label{fig:density}
\end{figure}

Let us comment on a possible cosmological scenario in which $\alpha_* \ll 1$ while the phase transition is strongly supercooled.
Such a situation can naturally occur when the bubble-nucleation sector remains completely decoupled from the visible Standard Model (SM) sector throughout the transition.
Denoting the plasma energy densities of the SM sector and the bubble-nucleation sector by
$\rho_{\mathrm{SM}}$ and
$\rho_{\rm b, rad}$, respectively, 
we consider the hierarchy 
$\rho_{\rm b, rad} \ll \rho_0 \ll \rho_{\rm SM}$ ($\simeq \rho_{\rm tot}$) as illustrated schematically in Fig.~\ref{fig:density}. 
The condition $\rho_0 \ll \rho_{\mathrm{SM}}$ ensures that the total energy density is dominated by SM radiation, yielding $\alpha_* \simeq \rho_0/\rho_{\mathrm{SM}} \ll 1$. This justifies neglecting the backreaction of bubble nucleation on the background metric.
Meanwhile, within the bubble-nucleation sector, the requirement $\rho_{\mathrm{b, rad}} \ll \rho_{0}$ implies that the transition is supercooled and that the latent heat is sufficiently large, leading to a prolonged phase transition.
See also footnote~\ref{footnote1}. 
Although this provides a concrete realization of the conditions required for our analysis, the calculation that follows does not rely on this specific scenario. We simply assume that the backreaction of bubble nucleation on the metric is negligible.

The stochastic GW spectrum can be calculated as (see Appendix)
\begin{align}
\Omega_{\rm GW}(\tau,k) 
&\equiv \frac{1}{\rho_{\rm tot}(\tau)} \frac{d\rho_{\rm GW}(\tau,k)}{d\ln k} \nonumber \\
&= \kappa^2(\tau_*) \alpha_*^2 \left(\frac{\mathcal{H}_*}{\tilde{\beta}}\right)^2 \lmk \frac{a^4 (\tau_*) \rho_{\rm tot}(\tau_*)}{a^4 (\tau) \rho_{\rm tot}(\tau)} \rmk
\Delta(k, \tilde{\beta}) \,,
\label{eq:Omega_Delta}
\end{align}
where the function $\Delta$ is given by%
\footnote{
In Refs.~\cite{Jinno:2016vai,Jinno:2017fby}, the integration range of $\tau$ is restricted to a finite interval in order to explicitly specify the time at which the bubble nucleation rate turns on and off.
In our case, the nucleation rate can be time dependent, and the finiteness of the active period can be implemented directly in the nucleation rate itself, for example by inserting a theta function.
With this treatment, the integration domain for $\tau$ in the formula of \eq{eq:Delta} can safely be taken from $0$ to $\infty$.
Moreover, as discussed around \eq{eq:nucleation-rate-exp}, the factor $P_2$ defined below ensures that the effective GW source remains finite.
}
\begin{align}
\Delta(k, \tilde{\beta}) = \frac{3}{4\pi^2}\frac{\tilde{\beta}^2k^3}{\kappa^2(\tau_*) \rho_0^2(\tau_*)}
\int_0^\infty d\tau_x
\int_0^\infty d\tau_y \;
\lmk \frac{a (\tau_x) a(\tau_y)}{a^2(\tau_*)} \rmk^3 \cos(k(\tau_x - \tau_y))\Pi (\tau_x,\tau_y,k) \,.
\label{eq:Delta}
\end{align}
The efficiency factor $\kappa$ quantifies the fraction of vacuum energy converted into the energy of the bubble wall, and may in general depend on time.
The function $\Pi$ can be expressed in terms of the energy-momentum tensor $T_{kl}$ as
\begin{align}
&a^2 (\tau_x) a^2(\tau_y) \Pi(\tau_x,\tau_y,k) = K_{ij,kl}(\hat{k})K_{ij,mn}(\hat{k})
\int d^3r \; e^{i \vec{k} \cdot \vec{r}} \langle T_{kl} T_{mn} \rangle (\tau_x,\tau_y,\vec{r}) \,,
\label{eq:Pi}
\end{align}
where the two-point correlation function of the energy-momentum tensor is defined by 
\begin{align}
\langle T_{kl} T_{mn} \rangle (\tau_x,\tau_y,\vec{r})
&\equiv \langle T_{kl}(\tau_x,\vec{x}) T_{mn}(\tau_y,\vec{y}) \rangle \,,
\label{eq:TT}
\end{align}
with $\vec{r} \equiv \vec{x} - \vec{y}$.
Here, $K_{ij,kl}$ 
denotes the transverse-traceless projection operator (see Eq.~\eqref{eq:kk}), and the hat notation $\hat{\bullet}$ indicates a unit vector in the direction of $\vec{\bullet}$.

The energy-momentum tensor is evaluated for bubble walls.
We consider a thin-shell bubble wall with an infinitesimal comoving width $l_B$.%
\footnote{
This width can in general be time-dependent, but our final expressions remain unchanged in the limit $l_B(\tau) \to 0$ for all $\tau$.
}
We ultimately take the limit $l_B \to 0$ in our expressions. 
To begin, let us consider the energy-momentum tensor of an uncollided bubble wall nucleated at $x_n \equiv (\tau_n, \vec{x}_n)$, which is given by
\begin{align}
T_{ij}^B(x ; x_n)
&= a^2(\tau_x) \rho_B(x; x_n)\widehat{(x - x_n)}_i\widehat{(x - x_n)}_j \,,
\label{eq:TB}
\end{align}
where $x \equiv (\tau_x,\vec{x})$ and $\rho_B(x;x_n)$ denotes the energy density at point $x$ due to the bubble wall nucleated at $x_n$. 
The source of the energy for the bubble wall arises from the false vacuum energy density multiplied by the volume swept out by the bubble per unit time:
\beq
 \frac{1}{a(\tau)}\frac{d E_B}{d \tau} = 4 \pi a^2(\tau) r_B^2 (\tau; \tau_n) \kappa(\tau) \rho_0(\tau) \,,
\eeq
where $r_B(\tau; \tau_n) = \tau - \tau_n$ is the comoving radius of the bubble. 
The energy density of the bubble wall is then given by $\rho_B(x; x_n) = E_B/(4 \pi a^3(\tau_x) r_B^2 (\tau_x; \tau_n) l_B)$ 
or more explicitly  
\begin{align}
\rho_B(x; x_n)
&=
\left\{
\begin{array}{cc}
\displaystyle 
\rho_B(\tau_x; \tau_n)
&
r_B(\tau_x) < |\vec{x} - \vec{x}_n| < r_B(\tau_x) + l_B \\
0
& 
{\rm otherwise}
\end{array}
\right.  \,,
\label{eq:rho}
\end{align}
where 
\beq
 \rho_B(\tau; \tau_n) \equiv \frac{1}{4\pi a^3(\tau) r_B^2(\tau; \tau_n) l_B}
\int_{\tau_n}^\tau d \tau'~4 \pi a^3(\tau') r_B^2 (\tau'; \tau_n) \kappa(\tau') \rho_0(\tau') \,.
\eeq

\begin{figure}
 \centering
 \includegraphics[width=0.45\hsize]{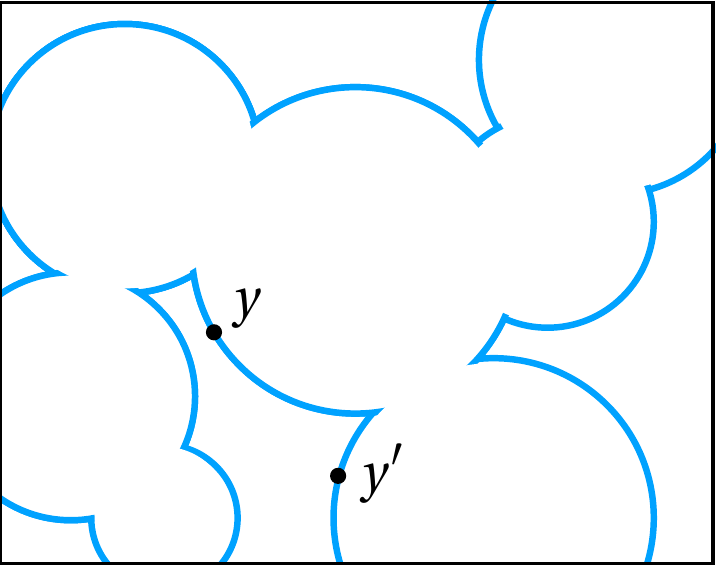}
 \quad 
 \includegraphics[width=0.45\hsize]{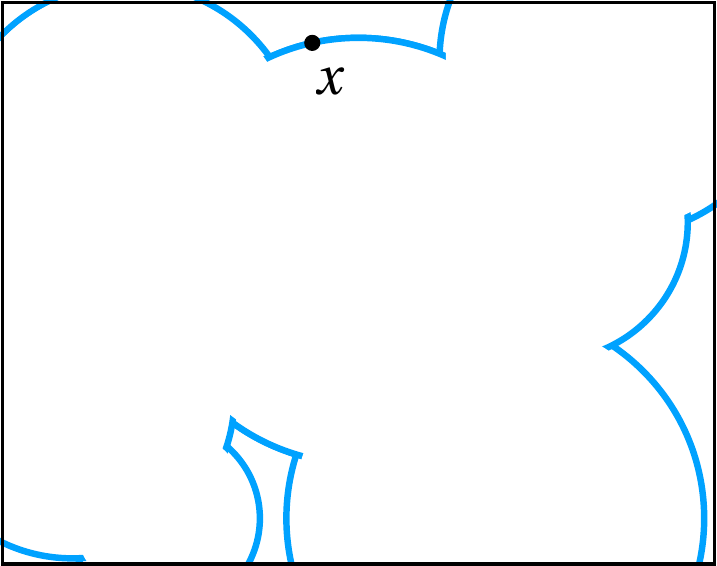}
 \caption{
 Schematic illustration of bubbles under the envelope approximation at different times. 
 Points $x$ and $y$ lie within the world volume of the same bubble, while point $y'$ lies within that of a different bubble. 
 The correlation function between $x$ and $y$ corresponds to the single-bubble contribution, whereas that between $x$ and $y'$ corresponds to the double-bubble contribution. 
}
 \label{fig:bubbles}
\end{figure}

The total energy-momentum tensor of the system becomes more intricate due to bubble collisions. In the following section, we analytically compute its correlation function under the envelope approximation. Figure~\ref{fig:bubbles} illustrates schematic configurations of bubbles at different times in this approximation. We may evaluate the correlation function between the points $x$ and $y$, or between $x$ and $y'$. The points $x$ and $y$ lie within the world volume of the same bubble, while $y'$ lies within that of a different bubble. The contribution from $x$ and $y$ is referred to as the single-bubble contribution, whereas that from $x$ and $y'$ is the double-bubble contribution. Note that the single-bubble contribution is also relevant (and indeed dominant) for the GW source, since even the region inside a single bubble loses spherical symmetry once collisions with other bubbles occur.

\section{Analytic derivation of GW spectrum in an expanding Universe}
\label{sec:analytic}

We now derive an analytic expression for the GW spectrum in an expanding Universe.
The calculations in this section largely parallel those presented in Refs.~\cite{Jinno:2016vai,Jinno:2017fby}, with the replacement of physical time by conformal time.
Accordingly, we omit detailed derivations and instead summarize the key steps and main results.

\subsection{Notations and definitions}

\begin{figure}
 \centering
 \includegraphics[width=0.5\hsize]{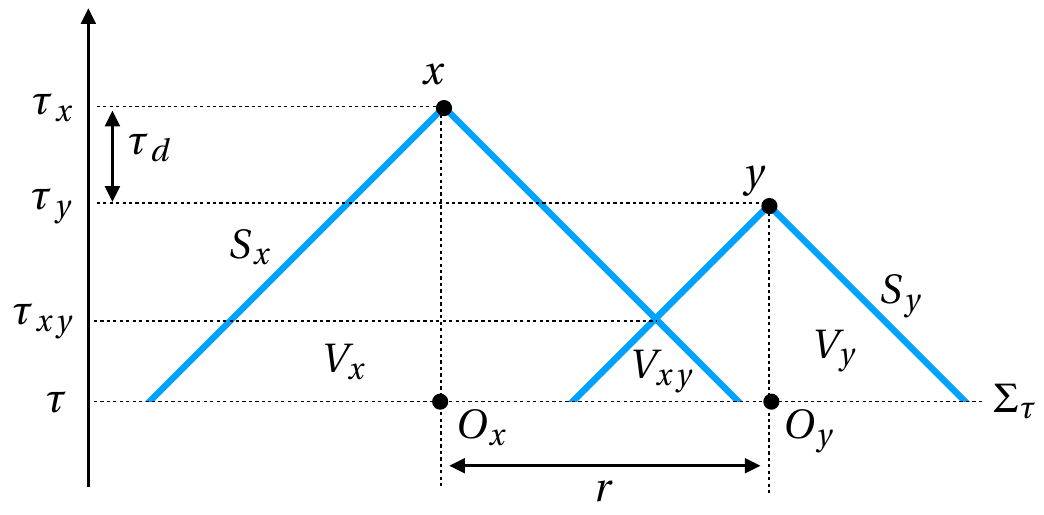}
 \quad 
 \includegraphics[width=0.4\hsize]{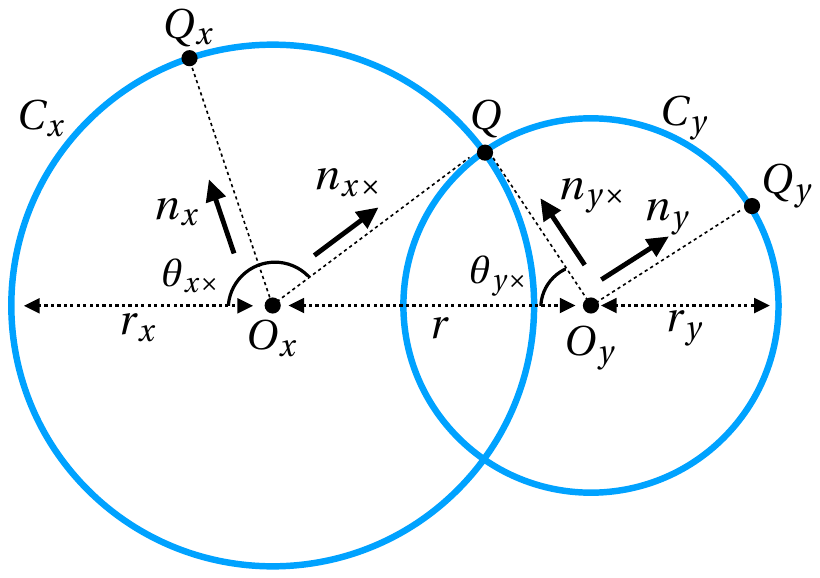}
 \caption{
 Schematic illustration of variables. 
 Left: causality of points $x$ and $y$ on a one-dimensional slice. 
 Right: unit vectors on a two-dimensional slice of a constant-time hypersurface $\Sigma_\tau$. 
}
 \label{fig:parameters}
\end{figure}

We begin by specifying the notations and conventions used throughout our calculations. See also Fig.~\ref{fig:parameters} for a schematic illustration of the relevant variables. These variables follow the original work, except that here we employ conformal variables. This choice is justified because the essential features, such as the causal structure and angular relations, remain unchanged under conformal rescaling.

We denote four-component comoving coordinates as
\begin{align}
x 
&= (\tau_x,\vec{x}) \,,
\;\;\;
y = (\tau_y,\vec{y}) \,.
\end{align}
We also define 
\begin{align}
x + \delta 
&\equiv (\tau_x + l_B,\vec{x}) \,,
\;\;\;
y + \delta \equiv (\tau_y + l_B,\vec{y}) \,,
\end{align}
and 
\begin{align}
\vec{r}
\equiv \vec{x} - \vec{y},
\;\;\;
r
&\equiv |\vec{r}| \,.
\end{align}
Hereafter, we use the index $i$ to generically denote any of the points $x$, $y$, $x+\delta$, and $y+\delta$. 
We further define
\begin{align}
&r_i(\tau)
\equiv 
\tau_i - \tau, 
\label{eq:ritau}
\\
&\tau_{xy}
\equiv \frac{\tau_x + \tau_y - r}{2} \,.
\end{align}
When performing integrals over $\tau_x$ and $\tau_y$, we change variables as
\begin{align}
&{\mathcal T}
\equiv \frac{\tau_x + \tau_y}{2}, 
\;\;\;
\tau_d
\equiv \tau_x - \tau_y
\quad
\iff
\quad
\tau_x = {\mathcal T} + \tau_d/2 \,,
\;\;\;
\tau_y = {\mathcal T} - \tau_d/2 \,.
\end{align}
We denote 
\beq
 &S_i \equiv ({\rm past \ light \ cone \ for \ point} \ i)  \,,
 \\
 &V_i \equiv ({\rm 4\text{-}volume \ region \ inside} \ S_i)  \,,
 \\
 &V_{xy} \equiv V_x \cup V_y \,. 
\eeq
Furthermore, 
\begin{align}
&\delta V_x
\equiv V_{x+\delta} - V_x \,,
\;\;\;
\delta V_y
\equiv V_{y+\delta} - V_y \,,
\\
&\delta V_{xy}
\equiv \delta V_x \cap \delta V_y \,,
\\
&\delta V_x^{(y)}
\equiv \delta V_x - V_{y+\delta} \,,
\;\;\;
\delta V_y^{(x)}
\equiv \delta V_y - V_{x+\delta} \,,
\end{align}
where $A-B$ denotes the region of the 4-volume $A$ that lies outside $B$.

Denoting a constant-time hypersurface at a conformal time $\tau$ by $\Sigma_\tau$, we define a sphere and its center as 
\beq
&C_i (\tau) \equiv S_x \cap \Sigma_\tau \,,
\\
&O_i \equiv {\rm center \ of} \ C_i(\tau) \,.
\eeq
Note that $r_i(\tau)$ defined in \eq{eq:ritau} represents the radius of $C_i(\tau)$.
Let $Q_i (\tau)$ be an arbitrary point on $C_i(\tau)$ (which is denoted as $P_i$ in Ref.~\cite{Jinno:2016vai}). We then define the unit vector
\beq
n_i (\tau) &\equiv {\rm unit \ vector \ from} \ O_i \ {\rm to} \ Q_i(\tau) 
\nonumber\\
&\equiv
(\sin \theta_i \cos \phi_i,\sin \theta_i \sin \phi_i,\cos \theta_i) \,,
\eeq
where in the second line, $n_i (\tau)$ is parametrized using the azimuthal and polar angles $(\phi_i, \theta_i)$ relative to the direction $\vec{r}$.

Next, let $Q(\tau)$ denote an arbitrary point on the intersection $C_x(\tau) \cap C_y(\tau)$ (which is denoted as $P$ in Ref.~\cite{Jinno:2016vai}). We define
\beq
n_{i\times} (\tau) &\equiv {\rm unit \ vector \ from} \ O_i \ {\rm to} \ Q(\tau)
\nonumber\\
&\equiv (\sin \theta_{i\times} \cos \phi_{i\times},\sin \theta_{i\times} \sin \phi_{i\times},\cos \theta_{i\times}) \,.
\eeq
In particular, the polar angles $\theta_{x\times}$ and $\theta_{y\times}$ satisfy 
\begin{align}
\cos\theta_{x\times}(\tau)
&=
- \frac{r^2 + r_x^2(\tau) - r_y^2(\tau)}{2 r r_x(\tau)},
\qquad
\cos\theta_{y\times}(\tau)
=
\frac{r^2 + r_y^2(\tau) - r_x^2(\tau)}{2 r r_y(\tau)} \,,
\label{eq:cos}
\end{align}
(see the right panel of Fig.~\ref{fig:parameters}.)

\subsection{False-vacuum probability}

Using the notation introduced above,
the probability of finding a spatial point $\vec{x}$ in the false vacuum at time $\tau_x$ is given by $P_1(x) = e^{-I_1(x)}$, where 
\beq
I_1(x)
&=
\int_{V_x} d^4 z~\tilde{\Gamma} (\tau)
=
\int_0^{\tau_x} d\tau~\frac{4\pi}{3} r_x^3 (\tau) \tilde{\Gamma} (\tau) \,.
\eeq
Due to translational symmetry,
the spatial dependence is trivial, and the result reduces to \eq{eq:I}, as expected.
Thus, we may simply write $P_1(x) = P_1(\tau) = e^{-I_1(\tau)}$.

The probability that a pair of spatial points $\vec{x}$ and $\vec{y}$ remain in the false vacuum at times $\tau_x$ and $\tau_y$, respectively, is given by $P_2(x,y) = e^{-I_2(x,y)}$ with $I_2(x,y) = \int_{V_{xy}} d^4z \; \tilde{\Gamma}(z)$. 
The integral over $V_{xy}$ can be evaluated by decomposing the integration domain into the regions $\tau < \tau_{xy}$ and $\tau \ge \tau_{xy}$.
For $\tau < \tau_{xy}$, the integration domain at fixed $\tau$ corresponds to the region enclosed by the blue circles shown in the right panel of Fig.~\ref{fig:parameters}.
For $\tau \ge \tau_{xy}$, the relevant domain reduces to the past light cone of $x$ and/or $y$, since in this regime $V_x$ and $V_y$ no longer overlap.
We thus obtain
\begin{align}
I_2(x,y)
&= \int_{V_{xy}} d^4z \; \tilde{\Gamma}(z)
\\
&= \int_{0}^{\tau_{xy}} d\tau \; \tilde{\Gamma}(\tau) \frac{\pi}{3}
\left[ r_x^3(\tau) 
(2 + \cos \theta_{x\times}(\tau))(1 - \cos \theta_{x\times} (\tau))^2
\right.
\nonumber \\
&\;\;\;\;\;\;\;\;\;\; \;\;\;\;\;\;\;\;\;\; \;\;\;\;\;\;\;\;\;\; \;\;\;\;\;\;\;\;\;\;
\left.
+ r_y^3(\tau) 
(2 - \cos \theta_{y\times}(\tau))(1 + \cos \theta_{y\times} (\tau))^2 \right]
\nonumber \\
&\;\;\;\;\;\;\;\;\;\;
+
\int_{\tau_{xy}}^{\tau_x} d\tau \; \frac{4\pi}{3} r_x^3(\tau) \tilde{\Gamma}(\tau)
+
\int_{\tau_{xy}}^{\tau_y} d\tau \; \frac{4\pi}{3} r_y^3(\tau) \tilde{\Gamma}(\tau)
\\
&=
 \int_{0}^{\tau_{xy}} d\tau \; \frac{\pi}{3} \tilde{\Gamma}(\tau) 
 \lkk 
 \frac{\lmk r + 2 ({\mathcal T} - \tau) \rmk^2}{4 r} 
 \lmk 3 \tau_d^2 - r^2 + 4 r ({\mathcal T} - \tau) \rmk 
 \rkk
 \nonumber\\ 
 &\;\;\;\;\;\;\;\;\;\;
 + \int_{\tau_{xy}}^{\tau_x} d\tau \; \frac{4\pi}{3} \tilde{\Gamma}(\tau) \lmk {\mathcal T} + \tau_d/2 -\tau \rmk^3 + \int_{\tau_{xy}}^{\tau_y} d\tau \; \frac{4\pi}{3} \tilde{\Gamma}(\tau) \lmk {\mathcal T} - \tau_d/2 -\tau \rmk^3 \,, 
 \label{eq:I2xy}
\end{align}
where we have used \eq{eq:cos}.%
\footnote{
Note that when $x$ and $y$ are exchanged, one simultaneously has $\vec{r} \to -\vec{r}$, implying that
$\cos\theta_{x\times} \leftrightarrow -\cos\theta_{y\times}$. 
}
This expression depends only on $\tau_x$, $\tau_y$, and $r$ (or equivalently on ${\mathcal T}$, $\tau_d$, and $r$), so that we denote it compactly as $P_2(x,y) = P_2(\tau_x, \tau_y,r) = e^{-I_2(\tau_x, \tau_y,r) }$. 
Note that we implicitly assume $\tau_{xy} \ge 0$, $\tau_{xy} \le \tau_x$, and $\tau_{xy} \le \tau_y$ in the above expression. 
These conditions can be rewritten as $\abs{\tau_d} \le r \le 2  {\mathcal T}$. 
This is sufficient for our purpose, since $P_2$ contributes only within this domain.

\subsection{Single-bubble spectrum}

According to Refs.~\cite{Jinno:2016vai,Jinno:2017fby}, 
the correlation of the energy-momentum tensor at points $x$ and $y$ consists of two distinct contributions. The first corresponds to the case in which both $x$ and $y$ lie on the world volume of a single bubble wall, realized when a bubble nucleates at a point $Q$ (see the right panel of Fig.~\ref{fig:parameters}), or more precisely, within the region $\delta V_{xy}$.
We denote quantities associated with this contribution using the superscript $(s)$.

The product of the energy-momentum tensors $T_{kl}(\tau_x,\vec{x}) T_{mn}(\tau_y,\vec{y})$ from the single-bubble contribution becomes nonzero only when both points $x$ and $y$ remain in the false vacuum, while the shifted points $x+\delta$ and $y+\delta$ enter the true vacuum due to the same single bubble.
Taking the ensemble average $\langle \cdot \rangle$, which is performed with respect to the stochastic nature of the bubble nucleation rate, the correlation of the energy–momentum tensor \eqref{eq:TT} is evaluated as 
\begin{align}
&\langle T_{kl} T_{mn} \rangle^{(s)} (\tau_x,\tau_y,\vec{r})
= 
P_2 (\tau_x, \tau_y, r) \int_{\delta V_{xy}} d^4 x_n 
\tilde{\Gamma}(\tau_n) T^B_{kl}(x;x_n) T^B_{mn}(y;x_n)\,,
\end{align}
where $T^B_{kl}(x;x_n)$ is given by \eq{eq:TB}. 
In the notation introduced above, $\widehat{(x - x_n)}_i$ appearing in \eq{eq:TB} can be written as $(-n_{x\times})_i$.
Decomposing the integration over $\delta V_{xy}$ into the integration over $\tau_n$ and over $\delta V_{xy} \cap \Sigma_{\tau_n}$, we obtain 
\begin{align}
&\frac{\langle T_{kl} T_{mn} \rangle^{(s)} (\tau_x,\tau_y,\vec{r})}{a^2(\tau_x) a^2(\tau_y)}
\nonumber \\
&= 
P_2 (\tau_x, \tau_y, r) \int_{0}^{\tau_{xy}} d\tau_n~
\tilde{\Gamma}(\tau_n) \rho_B(\tau_x ;\tau_n) \rho_B(\tau_y ;\tau_n) 
\int_{\delta V_{xy} \cap \Sigma_{\tau_n}} d^3z~(N_{\times}(\tau_n))_{klmn} \,,
\label{eq:single_TT}
\end{align}
where 
\begin{align}
(N_{\times} (\tau_n))_{klmn}
&\equiv
(n_{x\times})_k (n_{x\times})_l (n_{y\times})_m (n_{y\times})_n \,.
\end{align}
The integration domain $\delta V_{xy} \cap \Sigma_{\tau_n}$ corresponds to 
the intersection of the two circle boundaries in the right panel of Fig.~\ref{fig:parameters}, integrated over the azimuthal angle $\phi_{x\times}$ with ``radius" $r_x \sin \theta_{x \times}$.
The intersection of the two circle boundaries forms a diamond-shaped area near the point $Q$ for finite $l_B$, whose area is
$l_B^2 / \sin (\pi - \theta_{x\times} + \theta_{y\times}) = l_B^2  r_y / (r \sin \theta_{x \times})$. 
Thus we obtain 
\begin{align}
\int_{\delta V_{xy} \cap \Sigma_{\tau_n}} d^3z \; 
 (N_{\times}(\tau_n))_{klmn}
=
\frac{r_x r_y l_B^2}{r} \int_0^{2\pi} d \phi_{x\times} 
(N_{\times}(\tau_n))_{klmn} \,,
\end{align}
where we implicitly assume $\abs{\tau_x - \tau_y} < r < \tau_x + \tau_y$ 
so that $\delta V_{xy} \cap \Sigma_{\tau_n}$ is nonempty.

We now evaluate \eq{eq:Pi} by substituting \eq{eq:single_TT}.
The projection operator acts on $(N_{\times})_{klmn}$ such as 
\begin{align}
&\frac{r_x r_y l_B^2}{r}
\int_0^{2\pi} d \phi_{x\times} 
K_{ij,kl}(\hat{k})K_{ij,mn}(\hat{k}) (N_{\times}(\tau_n))_{klmn}
\nonumber\\
&=
\frac{\pi l_B^2}{r r_x r_y}
\biggl[
{\mathcal S}_0
+ \frac{1}{2}(1 - (\hat{r} \cdot \hat{k})^2) {\mathcal S}_1
+ \frac{1}{8}(1 - (\hat{r} \cdot \hat{k})^2)^2 {\mathcal S}_2 
\biggr] \,,
\label{eq:KKN}
\end{align}
with
\begin{align}
{\mathcal S}_0
&=
r_x^2 r_y^2s_{x\times}^2s_{y\times}^2, 
\label{eq:F0pp}
\\
{\mathcal S}_1
&= 
r_x r_y \left[ 
4c_{x\times}c_{y\times}(r_x^2s_{x\times}^2 + r_y^2s_{y\times}^2)
- 2r_xr_ys_{x\times}^2s_{y\times}^2 
\right] \,,
\label{eq:F1pp}
\\
{\mathcal S}_2
&=
r_x r_y
\bigl[
r_xr_y(19c_{x\times}^2 c_{y\times}^2 - 7(c_{x\times}^2 + c_{y\times}^2) + 3) - 8c_{x\times}c_{y\times}(r_x^2s_{x\times}^2 + r_y^2s_{y\times}^2)
\bigr] \,.
\label{eq:F2pp}
\end{align}
Here, we have adopted abbreviated notations for simplicity:
$r_x(\tau_n) \equiv r_x$, $r_y(\tau_n) \equiv r_y$, 
$\cos \theta_{x \times}(\tau_n) \equiv c_{x \times}$, 
$\sin \theta_{x \times}(\tau_n) \equiv s_{x \times}$, 
$\cos \theta_{y \times}(\tau_n) \equiv c_{y \times}$, 
and
$\sin \theta_{y \times}(\tau_n) \equiv s_{y \times}$. 
Since the $\hat{r}$ dependence enters only through $(\hat{r} \cdot \hat{k})^2$ and $(\hat{r} \cdot \hat{k})^4$ in \eq{eq:KKN},
we can evaluate the Fourier integrals in \eq{eq:Pi}
using
\beq
\int d^3 r~e^{i \vec{k} \cdot \vec{r}}
&=
\int_0^\infty dr~4 \pi r^2 j_0 (k r),
\label{eq:j0}
\\
\int d^3 r~e^{i \vec{k} \cdot \vec{r}} \frac{1}{2} (1 - (\hat{r} \cdot \hat{k})^2)
&=
\int_0^\infty dr~4 \pi r^2 \frac{j_1 (k r)}{k r},
\label{eq:j1}
\\
\int d^3 r~e^{i \vec{k} \cdot \vec{r}} \frac{1}{8} (1 - (\hat{r} \cdot \hat{k})^2)^2
&=
\int_0^\infty dr~4 \pi r^2 \frac{j_2 (k r)}{k^2 r^2} \,,
\label{eq:j2}
\eeq
where $j_i(x)$ are the spherical Bessel functions. Using these identities and noting $\abs{\tau_x - \tau_y} \le r \le \tau_x + \tau_y$, we finally obtain
\beq
\Pi^{(s)} (\tau_x,\tau_y,k) 
&= 4 \pi^2 \int_{\abs{\tau_x - \tau_y}}^{\tau_x + \tau_y} d r \, r P_2 (\tau_x, \tau_y, r) \int_{0}^{\tau_{xy}} d\tau_n~\tilde{\Gamma}(\tau_n) 
\nonumber \\
&~~~~
\times
\frac{l_B^2 \rho_B(\tau_x ;\tau_n) \rho_B(\tau_y ;\tau_n)}{ r_x (\tau_n) r_y (\tau_n)} 
\left[
j_0 (k r) {\mathcal S}_0 + \frac{j_1 (k r)}{k r} {\mathcal S}_1 + \frac{j_2 (k r)}{k^2 r^2} {\mathcal S}_2 
\right] \,,
\eeq
or equivalently, 
\beq
\Delta^{(s)} 
(k,\beta) &= 3 \tilde{\beta}^2 k^3
\int_{0}^\infty d\tau_x
\int_{0}^\infty d\tau_y \; 
\lmk \frac{a (\tau_x) a(\tau_y)}{a^2(\tau_*)} \rmk^3 \cos(k(\tau_x - \tau_y)) 
 \int_{\abs{\tau_x - \tau_y}}^{\tau_x + \tau_y} d r \, r P_2 (\tau_x, \tau_y, r) 
\nonumber\\
&~~~~\times
\int_{0}^{\tau_{xy}} d\tau_n~\tilde{\Gamma}(\tau_n) \frac{l_B^2 \rho_B (\tau_x ;\tau_n) \rho_B(\tau_y ;\tau_n)}{\kappa^2(\tau_*) \rho_0^2 (\tau_*) r_x (\tau_n) r_y (\tau_n)}
\left[
j_0 (k r) {\mathcal S}_0 + \frac{j_1 (k r)}{k r} {\mathcal S}_1 + \frac{j_2 (k r)}{k^2 r^2} {\mathcal S}_2
\right] \,,
\label{eq:Deltas}
\eeq
from \eq{eq:Delta},
where ${\mathcal S}_i$ are defined in Eqs.~(\ref{eq:F0pp}\,\text{-}\,\ref{eq:F2pp}) together with Eq.~(\ref{eq:cos}).

It may be convenient to rewrite the integral in terms of ${\mathcal T}$ and $\tau_d$, as 
\beq
 \int_{0}^{\infty} d\tau_x
\int_{0}^{\infty} d\tau_y 
\int_{\abs{\tau_x - \tau_y}}^{\tau_x + \tau_y} d r 
\to 2 
\int_{0}^{\infty} d{\mathcal T} \int_{0}^{2 \mathcal T} d\tau_d 
\int_{\tau_d}^{2 \mathcal T} d r  \,, 
\label{eq:integrationdomain}
\eeq
where we used the fact that the integrand is an even function of $\tau_d$.

\subsection{Double-bubble spectrum}

The other contribution arises when the points $x$ and $y$ lie on the world volumes of two different bubble walls, realized when bubbles nucleate at points $Q_x$ and $Q_y$ (see the right panel of Fig.~\ref{fig:parameters}), or more precisely, within the regions $\delta V_x^{(y)}$ and $\delta V_y^{(x)}$.
We denote the quantities associated with this contribution by the superscript $(d)$.

The product of the energy-momentum tensors $T_{kl}(\tau_x,\vec{x}) T_{mn}(\tau_y,\vec{y})$ from the double-bubble contribution becomes nonzero when both points $x$ and $y$ remain in the false vacuum, while the shifted points $x+\delta$ and $y+\delta$ enter the true vacuum due to two different bubbles.
Taking the ensemble average,
the correlation of the energy-momentum tensor is evaluated as
\begin{align}
\frac{\langle T_{kl} T_{mn} \rangle^{(d)} (\tau_x,\tau_y,\vec{r})}{a^2(\tau_x) a^2(\tau_y) } 
&= 
P_2 (\tau_x, \tau_y, r) 
\int_{\delta V_x^{(y)}} d^4 x_n
\tilde{\Gamma}(\tau_{xn})
\frac{T^B_{kl}(x; x_{n}) }{a^2(\tau_x)}
\int_{\delta V_y^{(x)}} d^4 y_n
\tilde{\Gamma}(\tau_{yn})
\frac{T^B_{mn}(y; y_{n}) }{a^2(\tau_y)}
\nonumber\\
&=
P_2 (\tau_x, \tau_y, r) 
\int_{0}^{\tau_{xy}} d\tau_{xn}
\tilde{\Gamma}(\tau_{xn})
\rho_B(\tau_x, \tau_{xn}) 
\int_{\delta V_x^{(y)} \cap \Sigma_{\tau_{xn}}} d^3\vec{x}_n~(n_x)_k (n_x)_l
\nonumber \\
&~~~~
\times
\int_{0}^{\tau_{xy}} d\tau_{yn}
\tilde{\Gamma}(\tau_{yn})
\rho_B(\tau_y, \tau_{yn}) 
\int_{\delta V_y^{(x)} \cap \Sigma_{\tau_{yn}}} d^3\vec{y}_n~(n_y)_m (n_y)_n  + \dots \,,
\label{eq:Double_TT}
\end{align}
where we denote $x_n \equiv (\tau_{xn}, \vec{x}_n)$ and $y_n \equiv (\tau_{yn}, \vec{y}_n)$ as the integration variables. 
Here, the dots represent the contribution from $\tau_{xn}, \tau_{yn} > \tau_{xy}$, which does not affect the GW emission since $\delta V_x^{(y)} \cap \Sigma_{\tau_{xn}}$ and $\delta V_y^{(x)} \cap \Sigma_{\tau_{yn}}$ are spherically symmetric in that regime and thus yield no transverse-traceless component.
For the same reason, we focus on the regime $\abs{\tau_x - \tau_y} < r < \tau_x + \tau_y$.
The integrals over $\vec{x}_n$ and $\vec{y}_n$ can be performed as follows:
\begin{align}
\int_{\delta V_x^{(y)} \cap \Sigma_{\tau_{xn}}} d^3\vec{x}_n~(n_x)_i (n_x)_j
&= 
\pi l_B r_x^2 \int_{-1}^{\cos\theta_{x\times}} d \cos \theta_x
\lkk
\sin^2 \theta_x \delta_{ij}
+ 
\lmk 3 \cos^2 \theta_x - 1 \rmk 
\hat{r}_i \hat{r}_j
\rkk
\nonumber \\
&=
\pi l_B r_x^2 
\lkk 
\lmk \frac{2}{3} + \cos \theta_{x\times} - \frac{1}{3} \cos^3 \theta_{x\times} \rmk \delta_{ij}
+ 
\lmk \cos^3 \theta_{x\times} - \cos \theta_{x\times} \rmk
\hat{r}_i \hat{r}_j
\rkk \,,
\\
\int_{\delta V_y^{(x)} \cap \Sigma_{\tau_{yn}}} d^3\vec{y}_n~(n_y)_i (n_y)_j
&= 
\pi l_B r_y^2 \int_{\cos\theta_{y\times}}^1 d \cos \theta_y
\lkk
\sin^2 \theta_y \delta_{ij}
+ 
\lmk 3 \cos^2 \theta_y - 1 \rmk 
\hat{r}_i \hat{r}_j
\rkk
\nonumber \\
&=
- \pi l_B r_y^2 
\lkk 
\lmk \frac{2}{3} + \cos \theta_{y\times} - \frac{1}{3} \cos^3 \theta_{y\times} \rmk \delta_{ij}
+ 
\lmk \cos^3 \theta_{y\times} - \cos \theta_{y\times} \rmk
\hat{r}_i \hat{r}_j \rkk \,.
\end{align}
Acting the projection operator in \eq{eq:Pi}, 
only the term proortional to $\hat{r}_i \hat{r}_j$ 
contributes to the transverse-traceless part of the energy-momentum tensor. 
We therefore obtain
\begin{align}
&K_{ij,kl}(\hat{k})K_{ij,mn}(\hat{k})
\frac{\langle T_{kl} T_{mn} \rangle^{(d)} (\tau_x,\tau_y,\vec{r})}{a^2(\tau_x) a^2(\tau_y)}
\nonumber\\
&=
\frac{1}{2} P_2 (\tau_x, \tau_y, r)
{\mathcal D}^{(d)}_x(\tau_x,\tau_y, r)
{\mathcal D}^{(d)}_y(\tau_x,\tau_y, r)
(1 - (\hat{r} \cdot \hat{k})^2)^2 \,,
\label{eq:KKTT_double}
\end{align}
where we define 
\beq
&{\mathcal D}^{(d)}_x (\tau_x,\tau_y, r)
\equiv
\int_{0}^{\tau_{xy}} d\tau_{xn}~\tilde{\Gamma} (\tau_{xn}) \, \pi r_x^2 l_B \rho_B(\tau_x, \tau_{xn})
\lmk \cos^3 \theta_{x\times} (\tau_{xn}) - \cos \theta_{x\times} (\tau_{xn}) \rmk \,,
\\
&{\mathcal D}^{(d)}_y (\tau_x,\tau_y, r)
\equiv
\int_{0}^{\tau_{xy}} d\tau_{yn}~\tilde{\Gamma} (\tau_{yn}) \, \pi r_y^2 l_B \rho_B(\tau_y, \tau_{yn})
\lmk \cos \theta_{y\times} (\tau_{yn}) - \cos^3 \theta_{y\times} (\tau_{yn}) \rmk \,.
\eeq
The Fourier transform in \eq{eq:Pi} can be carried out explicitly using Eq.~(\ref{eq:j2}), yielding the final result
\begin{align}
&\Pi^{(d)}(\tau_x,\tau_y,k) = 
16 \pi \int_{\abs{\tau_x - \tau_y}}^{\tau_x + \tau_y} dr~r^2 P_2 (\tau_x, \tau_y, r) 
{\mathcal D}^{(d)}_x(\tau_x,\tau_y, r)
{\mathcal D}^{(d)}_y(\tau_x,\tau_y, r)
\frac{j_2 (k r)}{k^2 r^2} \,,
\end{align}
or equivalently, 
\begin{align}
\Delta^{(d)}(k, \beta)
&=
\frac{12 \tilde{\beta}^2 k^3}{\pi}
\int_0^\infty d\tau_x
\int_0^\infty d\tau_y
\lmk \frac{a (\tau_x) a(\tau_y)}{a^2(\tau_*)} \rmk^3
\cos(k(\tau_x - \tau_y)) 
\nonumber\\
&~~~~
\times
\int_{\abs{\tau_x - \tau_y}}^{\tau_x + \tau_y} dr~r^2 P_2 (\tau_x, \tau_y, r)
\frac{{\mathcal D}^{(d)}_x(\tau_x,\tau_y, r)
{\mathcal D}^{(d)}_y(\tau_x,\tau_y, r)}{\kappa^2(\tau_*) \rho_0^2(\tau_*)} \frac{j_2 (k r)}{k^2 r^2} \,,
\label{eq:Deltad}
\end{align}
from \eq{eq:Delta}.
Equations (\ref{eq:Deltas}) and (\ref{eq:Deltad}) represent additive contributions to the GW spectrum. They provide analytic expressions for the GW spectrum in an expanding Universe, valid for general functions $\rho_0(\tau)$ and $a(\tau)$.
These expressions can be evaluated numerically, for example, using a Monte Carlo algorithm. The numerical results will be presented in Sec.~\ref{sec:numerical}.

In Sec.~\ref{sec:asymptotic} and \ref{sec:largebeta}, we further simplify these equations under certain limiting cases, such as the large- and small-$k$ limits and the Minkovski limit. In these regimes, the integrals can be evaluated more easily, either numerically or analytically.

\section{Asymptotic behaviors}
\label{sec:asymptotic}

In this section, we analyze the asymptotic behavior of Eqs.~(\ref{eq:Deltas}) and (\ref{eq:Deltad}) in the limits of large and small $k$.
The integrals appearing in these expressions are significantly simplified in these regimes due to the absence of oscillatory terms. While the resulting expressions are strictly valid in the asymptotic limits $k \gg \tilde{\beta}$ and $k \ll \tilde{\beta}$, we expect that the full behavior is smoothly interpolated around $k \sim \tilde{\beta}$. This suggests that these asymptotic results can also serve as useful approximations for understanding the general dependence of the overall GW amplitude on $\tilde{\beta}/\mathcal{H}_*$.

\subsection{Large $k$ limit}
\label{sec:largek}
For $k \gg \tau_*^{-1}$, 
the integral over $r$ is dominated by the region $r \lesssim 1/k \ll \tau_*$, since the Bessel function becomes highly oscillatory for $kr \gg 1$. 
We also note that $\abs{\tau_d}$ ($\le r$) is similarly small. 
To capture the behavior in this regime,
we rescale the small quantities by substituting 
$r \to \epsilon r$ and $\tau_d \to \epsilon \tau_d$, and expand the relevant expressions in powers of the small parameter $\epsilon$. 
We set $\epsilon = 1$ at the end of the calculation.

Furthermore, the phase transition occurs within a (conformal) time scale of order $1/\tilde{\beta}$. 
This indicates that the integral over $\tau_n$ is dominated near 
$({\mathcal T} - \tau_n) \sim 1/\tilde{\beta}$. 
The integrand can therefore be simplified under the condition $\epsilon r \ll ({\mathcal T} - \tau_n) \sim 1/\tilde{\beta}$, which corresponds to the limit of $k \gg \tilde{\beta}$.

We eventually find that, in the expansion of the probability $P_2(\tau_x,\tau_y, r)$
\beq
P_2(\tau_x,\tau_y, r) = e^{-I_2({\mathcal T},{\mathcal T},0)} 
\lkk
1 - r \lmk 1 + \frac{\tau_d^2}{r^2} \rmk \int_{0}^{{\mathcal T}} d\tau \; \pi ({\mathcal T} - \tau)^2 \tilde{\Gamma}(\tau) \epsilon + \mathcal{O}(\epsilon^2)
\rkk \,,
\eeq
with 
\beq
I_2({\mathcal T},{\mathcal T},0) 
= 
\int_{0}^{{\mathcal T}} d\tau \; \frac{4 \pi}{3} ({\mathcal T} - \tau)^3 \tilde{\Gamma}(\tau) \,.
\eeq
The ${\mathcal O} (\epsilon)$
term provides the leading contribution at this order to both the single- and double-bubble spectra.

\subsubsection{Single-bubble contribution}

We begin with the single-bubble spectrum:
\beq
\Delta^{(s)} 
(k,\beta) &= 3 \tilde{\beta}^2 k^3
\int_{0}^\infty d\tau_x
\int_{0}^\infty d\tau_y \; 
\lmk \frac{a (\tau_x) a(\tau_y)}{a^2(\tau_*)} \rmk^3 \cos(k(\tau_x - \tau_y)) 
 \int_{\abs{\tau_x - \tau_y}}^{\tau_x + \tau_y} d r \, r P_2 (\tau_x, \tau_y, r) 
\nonumber\\
&~~~~\times
\int_{0}^{\tau_{xy}} d\tau_n~\tilde{\Gamma}(\tau_n) \frac{l_B^2 \rho_B (\tau_x ;\tau_n) \rho_B(\tau_y ;\tau_n)}{\kappa^2 (\tau_*) \rho_0^2 (\tau_*) r_x (\tau_n) r_y (\tau_n)}
\left[
j_0 (k r) {\mathcal S}_0 + \frac{j_1 (k r)}{k r} {\mathcal S}_1 + \frac{j_2 (k r)}{k^2 r^2} {\mathcal S}_2
\right] \,.
\label{eq:Deltas_App}
\eeq
The cosine terms in ${\mathcal S}_0, {\mathcal S}_1, {\mathcal S}_2$ can be expanded as
\begin{align}
\cos\theta_{x\times}(\tau_n)
&= - \frac{2 \tau_d ({\mathcal T} - \tau_n) + r^2}{2({\mathcal T} +\tau_d/2 - \tau_n) r} 
= - \frac{\tau_d}{r} - \frac{r - \tau_d^2/r}{2({\mathcal T} - \tau_n)}\epsilon + \mathcal{O}(\epsilon^2) ,
\\
\cos\theta_{y\times}(\tau_n)
&= - \frac{2 \tau_d ({\mathcal T} - \tau_n) - r^2}{2({\mathcal T} -\tau_d/2 - \tau_n) r} 
= - \frac{\tau_d}{r} + \frac{r - \tau_d^2/r}{2({\mathcal T} - \tau_n)}\epsilon + \mathcal{O}(\epsilon^2) \,,
\end{align}
which yields
\begin{align}
{\mathcal S}_0 (\tau_x,\tau_y,\tau_n,r)
&=
({\mathcal T} - \tau_n)^4 \lmk 1- \frac{\tau_d^2}{r^2} \rmk^2 + \mathcal{O}(\epsilon^2),
\\
{\mathcal S}_1 (\tau_x,\tau_y,\tau_n,r)
&= 
({\mathcal T} - \tau_n)^4 \left[ 
- 10 \lmk \frac{\tau_d^2}{r^2} \rmk^2 + 12 \lmk \frac{\tau_d^2}{r^2} \rmk 
- 2 
\right] + \mathcal{O}(\epsilon^2),
\\
{\mathcal S}_2 (\tau_x,\tau_y,\tau_n,r)
&=
({\mathcal T} - \tau_n)^4 \lkk 35 \lmk \frac{\tau_d^2}{r^2} \rmk^2 - 30 \frac{\tau_d^2}{r^2} + 3 
\rkk + \mathcal{O}(\epsilon^2) \,.
\label{eq:Sseries}
\end{align}
Additionally, we have
\beq
\rho_B(\tau_x;\tau_n) \rho_B(\tau_y;\tau_n)
&= 
\rho_B^2({\mathcal T};\tau_n) + \mathcal{O}(\epsilon^2),
\\
r_x(\tau_n) r_y(\tau_n)
&=
({\mathcal T} - \tau_n)^2 + \mathcal{O}(\epsilon^2),
\\
a (\tau_x) a(\tau_y)
&=
a^2 ({\mathcal T}) + \mathcal{O}(\epsilon^2) \,.
\eeq
Note that the leading term of the integrand in Eq.~(\ref{eq:Deltas_App}) vanishes when $\tau_n = \tau_{xy} \simeq {\mathcal T}$.
Therefore, we may set the upper limit of the $\tau_n$ integration to ${\mathcal T}$, as long as we are interested only in contributions of order ${\mathcal O}(\epsilon)$.

In the large-$k$ limit, it is convenient to change the order of integration as
\beq
\int_{0}^\infty d\tau_x
\int_{0}^\infty d\tau_y 
\int_{\abs{\tau_x - \tau_y}}^{\tau_x + \tau_y} d r
\to
\int_0^\infty d{\mathcal T} 
\int_0^{2{\mathcal T} } d r 
\int_{-r}^r d\tau_d \,.
\label{eq:integrationdomain2}
\eeq
We then approximately extend the upper bound of the $r$ integral from $2 {\mathcal T}$ to $\infty$, which is justified when the integral is dominated by the regime $2 {\mathcal T} \gg r$. Under this assumption, the integrals over $r$ and $\tau_d$ can be carried out analytically.
Using these expressions, we can approximate the single-bubble spectrum as
\beq
\Delta^{(s)} 
(k/\tilde{\beta} \gg 1)
&\simeq
3 \tilde{\beta}^2 k^3
\int_0^\infty d{\mathcal T} 
\int_0^\infty d r \, r 
\int_{-r}^r d\tau_d~
\cos(k\tau_d) \lmk \frac{a ({\mathcal T})}{a(\tau_*)} \rmk^6 
P_2 (\tau_x, \tau_y, r) 
\nonumber \\
&~~~~
\times
\int_{0}^{\mathcal T} d\tau_n~\tilde{\Gamma}(\tau_n) 
\frac{l_B^2 \rho_B({\mathcal T}; \tau_n) \rho_B({\mathcal T}; \tau_n)}{({\mathcal T} - \tau_n)^2 \kappa^2(\tau_*) \rho_0^2 (\tau_*) } 
\left[
j_0 (k r) {\mathcal S}_0 + \frac{j_1 (k r)}{k r} {\mathcal S}_1 + \frac{j_2 (k r)}{k^2r^2} {\mathcal S}_2
\right]
\nonumber \\
&\simeq
\frac{4 \pi \tilde{\beta}^2}{k}
\int_0^\infty d{\mathcal T} \; 
e^{-I_2({\mathcal T},{\mathcal T},0)} \lmk \frac{a ({\mathcal T})}{a(\tau_*)} \rmk^6 
\int_0^{\mathcal T} d \tau ({\mathcal T} - \tau)^2 \tilde{\Gamma}(\tau)
\nonumber\\
&~~~~
\times
\int_{0}^{{\mathcal T}} d\tau_n~\tilde{\Gamma}(\tau_n) ({\mathcal T} - \tau_n)^2 \frac{l_B^2 \rho_B({\mathcal T}; \tau_n) \rho_B({\mathcal T}; \tau_n)}{\kappa^2 (\tau_*) \rho_0^2(\tau_*)} \epsilon + \mathcal{O}(\epsilon^2) \,.
\label{eq:largekresult}
\eeq
In the second line, the $\mathcal{O}(\epsilon^0)$ terms vanish upon integration over $r$ and $\tau_d$.
The resulting expression is proportional to $k^{-1}$, consistent with expectations from the literature.
Moreover, the integral in \eq{eq:largekresult} contains no oscillatory components, making it straightforward to evaluate numerically once the functional forms of $\tilde{\Gamma}(\tau)$, $a(\tau)$, and $\rho_0(\tau)$ are specified.

\subsubsection{Double-bubble contribution}

We now turn to the double-bubble contribution:
\begin{align}
\Delta^{(d)}(k, \beta)
&=
\frac{12 \tilde{\beta}^2 k^3}{\pi}
\int_0^\infty d\tau_x
\int_0^\infty d\tau_y
\lmk \frac{a (\tau_x) a(\tau_y)}{a^2(\tau_*)} \rmk^3
\cos(k(\tau_x - \tau_y)) 
\nonumber\\
&~~~~
\times
\int_{\abs{\tau_x - \tau_y}}^{\tau_x + \tau_y} dr~r^2 P_2 (\tau_x, \tau_y, r)
\frac{{\mathcal D}^{(d)}_x(\tau_x,\tau_y, r)
{\mathcal D}^{(d)}_y(\tau_x,\tau_y, r)}{\kappa^2 (\tau_*) \rho_0^2(\tau_*)} \frac{j_2 (k r)}{k^2 r^2} \,.
\label{eq:Deltad_App}
\end{align}
By decomposing the cosine terms, we obtain
\beq
&{\mathcal D}^{(d)}_x (\tau_x,\tau_y, r) {\mathcal D}^{(d)}_y (\tau_x,\tau_y, r)
\nonumber \\
&=
\pi^2 l_B^2
\int_0^{{\mathcal T} - \epsilon r / 2} d\tau_{nx}
\int_0^{{\mathcal T} - \epsilon r / 2} d\tau_{ny}~
\rho_B \lmk {\mathcal T} + \frac{\epsilon \tau_d}{2}, \tau_{xn} \rmk \rho_B \lmk {\mathcal T} - \frac{\epsilon \tau_d}{2}, \tau_{yn} \rmk
\nonumber \\
&~~~~
\times
\lkk
- \frac{\tau_d^2 (\tau_d^2 - r^2)^2}{r^6} ({\mathcal T} - \tau_{nx})^2 ({\mathcal T} - \tau_{ny})^2
-
\frac{\tau_d (\tau_d^2 - r^2)^3 (\tau_{nx} - \tau_{ny})}{2 r^6} ({\mathcal T} - \tau_{nx}) ({\mathcal T} - \tau_{ny}) \epsilon
\rkk 
\nonumber\\
&~~~~ + {\mathcal O} (\epsilon^2)
\,.
\label{eq:DDapprox}
\eeq
When we express the integral over $\tau_{nx}$ as $\int_0^{\mathcal{T}} d \tau_{nx} - \int_{\mathcal{T} - \epsilon r/2}^{\mathcal{T}} d \tau_{nx}$, the latter integral does not contribute to the overall ${\mathcal O} (\epsilon^1)$ correction, because the ${\mathcal O} (\epsilon^0)$ term inside the square brackets in \eq{eq:DDapprox} contains a factor of $({\mathcal T} - \tau_{nx})^2 = \mathcal{O}(\epsilon^2)$. Similarly, the integral over $\tau_{ny}$ can be written as $\int_0^{\mathcal{T}} d \tau_{nx}$ up to corrections of $\mathcal{O}(\epsilon^2)$.
The $\rho_B \lmk {\mathcal T} + \frac{\epsilon \tau_d}{2}, \tau_{xn} \rmk \rho_B \lmk {\mathcal T} - \frac{\epsilon \tau_d}{2}, \tau_{yn} \rmk$ temrs also do not contribute to the overall ${\mathcal O} (\epsilon^1)$ correction because
\beq
\int_0^{{\mathcal T}} d\tau_{nx}
\int_0^{{\mathcal T}} d\tau_{ny}~
&
\lkk \partial_{\mathcal T} \rho_B ({\mathcal T}, \tau_{xn}) \cdot \rho_B ({\mathcal T}, \tau_{yn}) - \rho_B ({\mathcal T}, \tau_{xn}) \cdot \partial_{\mathcal T} \rho_B ({\mathcal T}, 
\tau_{yn}) \rkk
\nonumber \\
&\times
({\rm symmetric~terms~under~}\tau_{xn} \leftrightarrow \tau_{yn})
=
0 \,.
\eeq
Moreover, the ${\mathcal O} (\epsilon^1)$ term in the square brackets in \eq{eq:DDapprox} also does not contribute to the total ${\mathcal O} (\epsilon^1)$ term, owing to
\beq
\int_0^{{\mathcal T}} d\tau_{nx}
\int_0^{{\mathcal T}} d\tau_{ny}~
&
(\tau_{nx} - \tau_{ny}) \times ({\rm symmetric~terms~under~}\tau_{xn} \leftrightarrow \tau_{yn})
=
0 \,.
\eeq
Thus, we arrive at
\beq
&{\mathcal D}^{(d)}_x (\tau_x,\tau_y, r) {\mathcal D}^{(d)}_y (\tau_x,\tau_y, r)
\nonumber \\
&=
-
\pi^2 l_B^2
\int_0^{\mathcal T} d\tau_{nx}
\int_0^{\mathcal T} d\tau_{ny}~
\rho_B ({\mathcal T}, \tau_{xn}) \rho_B ({\mathcal T}, \tau_{yn})
\frac{\tau_d^2 (\tau_d^2 - r^2)^2}{r^6} ({\mathcal T} - \tau_{nx})^2 ({\mathcal T} - \tau_{ny})^2
+
{\mathcal O} (\epsilon^2) \,.
\eeq
Therefore,
\begin{align}
\Delta^{(d)} (k/\tilde{\beta} \gg 1)
&=
\frac{12 \tilde{\beta}^2 k^3}{\pi}
\int_0^\infty d\tau_x
\int_0^\infty d\tau_y
\lmk \frac{a ({\mathcal T})}{a (\tau_*)} \rmk^6
\cos(k(\tau_x - \tau_y)) 
\nonumber\\
&~~~~
\times
\int_{\abs{\tau_x - \tau_y}}^{\tau_x + \tau_y} dr~r^2 P_2 (\tau_x, \tau_y, r)
\frac{{\mathcal D}^{(d)}_x(\tau_x,\tau_y, r)
{\mathcal D}^{(d)}_y(\tau_x,\tau_y, r)}{\kappa^2(\tau_*)\rho_0^2(\tau_*)} \frac{j_2 (k r)}{k^2 r^2}
+ {\mathcal O} (\epsilon^2)
\nonumber \\
&=
- 12 \pi \tilde{\beta}^2k^3
\int_0^\infty d {\mathcal T}
\int_0^\infty d r \, r^2
\int_{-r}^r d\tau_d \;
\lmk \frac{a ({\mathcal T})}{a(\tau_*)} \rmk^6 
\cos(k\tau_d)
P_2 (\tau_x, \tau_y, r)
\nonumber \\
&~~~~
\times
\lkk \int_{0}^{{\mathcal T}} d\tau_{n}~\tilde{\Gamma} (\tau_{n}) ({\mathcal T} - \tau_{n})^2 \lmk \frac{l_B \rho_B({\mathcal T}, \tau_{n})}{\kappa(\tau_*) \rho_0(\tau_*)}
\rmk \rkk^2 
\frac{\tau_d^2 (\tau_d^2 - r^2)^2}{r^6}
\frac{j_2 (k r)}{k^2r^2}
+ {\mathcal O} (\epsilon^2) \,,
\end{align}
where we change the integration domain as in \eq{eq:integrationdomain2} and then approximately extend the integration range of $r$ according to the discussion below \eq{eq:integrationdomain2}.
The leading ${\mathcal O} (\epsilon^0)$ contribution cancels once again, making the ${\mathcal O} (\epsilon)$ term in $P_2$ the dominant contribution such as 
\begin{align}
\Delta^{(d)} (k/\tilde{\beta} \gg 1)
&=
\frac{128\pi^3 \tilde{\beta}^2}{35k^2} 
\int_0^\infty d {\mathcal T} \;
e^{-I_2({\mathcal T},{\mathcal T},0)} \lmk \frac{a ({\mathcal T})}{a(\tau_*)} \rmk^6 
\int_0^{\mathcal T} d \tau ({\mathcal T} - \tau)^2 \tilde{\Gamma}(\tau)
\nonumber \\
&~~~~
\times 
\lkk \int_{0}^{{\mathcal T}} d\tau_{n}~\tilde{\Gamma} (\tau_{n}) 
({\mathcal T} - \tau_{n})^2 \lmk \frac{l_B \rho_B({\mathcal T}, \tau_{n}) }{\kappa (\tau_*)\rho_0(\tau_*)} \rmk
\rkk^2 \epsilon + \mathcal{O}(\epsilon^2) \,.
\label{eq:largekresultdouble}
\end{align}
This term scales as $k^{-2}$, in agreement with results from Ref.~\cite{Jinno:2016vai}.

\subsection{Small $k$ limit}
\label{sec:smallk}

We now consider the opposite limit, where $k$ is sufficiently small. 
As before, the integrals over $\tau_d$ and $r$ are typically dominated by the region $\lesssim 1/\tilde{\beta}$. 
For $k \ll \tilde{\beta}$, 
we can approximate $\cos (k \tau_d) \approx 1$ 
and $j_n(kr) / (kr)^n \approx \sqrt{\pi}/(2^{n+1}\Gamma(n+3/2))$. 

\subsubsection{Single-bubble contribution}

For the single-bubble spectrum, we obtain
\beq
\Delta^{(s)} 
(k/\tilde{\beta} \ll 1)
&\simeq
3 \tilde{\beta}^2 k^3
\int_0^\infty d\tau_x
\int_0^\infty d\tau_y
\lmk \frac{a (\tau_x) a(\tau_y)}{a^2(\tau_*)} \rmk^3
\int_{\abs{\tau_x - \tau_y}}^{\tau_x + \tau_y} d r~r P_2 (\tau_x, \tau_y, r) 
\nonumber\\
&~~~~
\times
\int_{0}^{\tau_{xy}} d\tau_n~\tilde{\Gamma}(\tau_n) \frac{l_B^2 \rho_B(\tau_x ;\tau_n) \rho_B(\tau_y; \tau_n)}{r_x (\tau_n) r_y (\tau_n) \kappa^2 (\tau_*) \rho_0^2(\tau_*) } 
\left[
{\mathcal S}_0 + \frac{1}{3}{\mathcal S}_1 + \frac{1}{15}{\mathcal S}_2
\right] \,.
\label{eq:smallkresult}
\eeq
Here, 
\beq
&{\mathcal S}_0 + \frac{1}{3}{\mathcal S}_1 + \frac{1}{15}{\mathcal S}_2 
= 
\frac{8}{15} \lmk {\mathcal T} - \tau_n \rmk^4
+ \frac{4}{15} \lmk {\mathcal T} - \tau_n \rmk^2 \lmk 2 \tau_d^2 - 3 r^2 \rmk
+ \frac{1}{30} \lmk \tau_d^4 - 6 \tau_d^2 r^2 + 6 r^4 \rmk \,.
\eeq
Since the integrand does not oscillate, the integrals can be easily evaluated numerically once the functional forms $a(\tau)$, $\rho_0(\tau)$, and $\tilde{\Gamma}(\tau)$ are specified. 
As expected from the literature, the result scales as $k^3$.

\subsubsection{Double-bubble contribution}

For the double-bubble contribution, we obtain 
\begin{align}
\Delta^{(d)}(k/\tilde{\beta} \ll 1)
&=
\frac{12 \tilde{\beta}^2 k^3}{\pi}
\int_0^\infty d\tau_x
\int_0^\infty d\tau_y
\lmk \frac{a (\tau_x) a(\tau_y)}{a^2(\tau_*)} \rmk^3
\int_{\abs{\tau_x - \tau_y}}^{\tau_x + \tau_y} dr~r^2 P_2 (\tau_x, \tau_y, r)
\nonumber\\
&~~~~
\times
\frac{{\mathcal D}^{(d)}_x(\tau_x,\tau_y, r) {\mathcal D}^{(d)}_y(\tau_x,\tau_y, r)}{\kappa^2(\tau_*) \rho_0^2(\tau_*)} \frac{1}{15} \,.
\end{align}
As expected from the literature, the result scales as $k^3$. 
Here, although the following expressions are not simplified within this limit, we explicitly present them for the reader's convenience:
\beq
&{\mathcal D}^{(d)}_x (\tau_x,\tau_y, r)
=
-\int_{0}^{\tau_{xy}} d\tau_{xn}~\tilde{\Gamma} (\tau_{xn}) \, \pi l_B \rho_B(\tau_x, \tau_{xn})
\nonumber\\
&\qquad \qquad \qquad  \qquad \qquad  \times
\frac{\lmk r^2 - \tau_d^2 \rmk \lmk r^2 + 2 \tau_d ( {\mathcal T} - \tau_{xn}) \rmk \lmk r^2 - 4 ( {\mathcal T} - \tau_{xn})^2 \rmk}{4 r^3 (2 {\mathcal T} + \tau_d - 2 \tau_{xn})}, 
\\
&{\mathcal D}^{(d)}_y (\tau_x,\tau_y, r)
=
-\int_{0}^{\tau_{xy}} d\tau_{yn}~\tilde{\Gamma} (\tau_{yn}) \, \pi l_B \rho_B(\tau_y, \tau_{yn})
\nonumber\\
&\qquad \qquad \qquad  \qquad \qquad  \times
\frac{\lmk r^2 - \tau_d^2 \rmk \lmk r^2 - 2 \tau_d ( {\mathcal T} - \tau_{yn}) \rmk \lmk r^2 - 4 ( {\mathcal T} - \tau_{yn})^2 \rmk}{4 r^3 (2 {\mathcal T} - \tau_d - 2 \tau_{yn})} \,.
\eeq

Finally, 
let us emphasize that the expression in this section does not contain oscillatory integrals, making it numerically tractable once the functional forms of $\tilde{\Gamma}(\tau)$, $a(\tau)$, and $\rho_0(\tau)$ are specified.

\section{Large $\tilde{\beta}/\mathcal{H}_*$ limit (expansion around the Minkovski limit)}
\label{sec:largebeta}

In this section, we derive analytic formulas in the large $\tilde{\beta}/\mathcal{H}_*$ limit, corresponding to the Minkovski approximation.

When the nucleation rate sharply increases around a specific time $\tau_*$, the phase transition completes within a time scale of order $1/\tilde{\beta}$. A large value of $\tilde{\beta}/\mathcal{H}_*$ thus implies that the transition occurs rapidly compared to the Hubble time, rendering the effect of cosmic expansion negligible. Therefore, this limit can be interpreted as the Minkovski limit.
We expand Eqs.~(\ref{eq:Deltas}) and (\ref{eq:Deltad}) in this limit and present analytic expressions for several representative cases.

To illustrate which terms can be simplified, let us for a moment focus on the specific case 
$\tilde{\Gamma} = \tilde{\Gamma}_0 e^{\tilde{\beta}' \tau }$ with $\tilde{\beta}/\mathcal{H}_* \gg 1$, where $\tilde{\Gamma}_0$ and $\tilde{\beta}'$ are constants. Note first that $\tilde{\beta}' \simeq \tilde{\beta}$ and $\log ( \tilde{\beta}'^4 / (8\pi \tilde{\Gamma}_0)) \simeq \tilde{\beta} /\mathcal{H}_* = \tilde{\beta} \tau_*$ in this case (see Eq.~(\ref{eq:approxbetatau})). 
From \eq{eq:Deltas}, 
the integral over $\tau_n$ is dominated near the upper boundary of the domain, $\tau_n \simeq \tau_{xy} = {\mathcal T} - r/2$, 
due to the exponential dependence introduced by $\tilde{\Gamma}$. 
On the other hand,
$P_2(\tau_x, \tau_y, r)$ is strongly suppressed when $\tilde{\beta} {\mathcal T} \gtrsim \log \lmk 8 \pi \tilde{\Gamma}_0 / \tilde{\beta}^4 \rmk$ ($\simeq \tilde{\beta} \tau_*$). 
Combining these observations,
the integral over ${\mathcal T}$ is dominated around ${\mathcal T} \sim \tau_*$ 
and the relevant range of $r$ is 
$\abs{\tau_d} \le r \lesssim 1/\tilde{\beta}$. 
Assuming $\tilde{\beta}/\mathcal{H}_* \gg 1$ (or $\tilde{\Gamma}_0 / \tilde{\beta}^4 \ll 1$), 
we can expand the integrand in terms of $\abs{\tau_d} /{\mathcal T}$ and $r/ {\mathcal T}$. 
This is analogous to the large-$k$ limit, except that ${\mathcal S}_i$ cannot be expanded as in \eq{eq:Sseries}, since ${\mathcal T} - \tau_n$ can also be a small quantity in the present case.
In particular, we can approximate $a(\tau_x) \approx a({\mathcal T}) \approx a(\tau_*)$ at leading order. 
Moreover, the integral over conformal time can be mathematically extended to $-\infty$ instead of starting from $0$ because the nucleation rate is exponentially suppressed in the extended domain.
At leading order, this allows us to reproduce the results of Ref.~\cite{Jinno:2016vai}, as we explain in Sec.~\ref{sec:analytic-exp} in more detail. 
A similar approximation generally applies to other forms of $\tilde{\Gamma}(\tau)$ as well.

To calculate the next-to-leading order term, we use 
\beq
a (\tau_x) a(\tau_y) 
&= a ({\mathcal T} + \tau_d/2) a({\mathcal T} - \tau_d/2) 
\simeq a^2 ({\mathcal T}) 
\\
&\simeq a^2(\tau_*) \lkk 1 + 2 ({\mathcal T} - \tau_*) \frac{a'(\tau_*)}{a(\tau_*)} \rkk 
\\
&= a^2(\tau_*) \lkk 1 + \frac{1}{3} (\tau_* - {\mathcal T}) \tilde{\beta} c_{\rm NL1} \rkk \,,
\eeq
where we define 
\beq
 &c_{\rm NL1} \equiv - \frac{6}{\tilde{\beta}} \frac{a'(\tau_*)}{a(\tau_*)} \,.
 \label{eq:CLNs2}
\eeq
We also have 
\beq
\frac{l_B \rho_B(\tau_x ;\tau_n) }{r_x (\tau_n) \kappa \rho_0(\tau) } \simeq \frac{1}{3}\lkk 1 - \frac{1}{4} \lmk \tau_x-\tau_n \rmk \lmk 3 \frac{a'(\tau_x)}{a(\tau_x)} + \frac{\rho_0'(\tau_x)}{\rho_0(\tau_x)} \rmk \rkk \,,
\eeq
for $(\tau_x - \tau_n) \ll \tau_x$. This gives 
\beq
\frac{l_B^2 \rho_B(\tau_x ;\tau_n) \rho_B(\tau_y ;\tau_n)}{r_x (\tau_n) r_y (\tau_n) \kappa^2 \rho_0^2({\mathcal T}) } 
&\simeq \frac{1}{9} \lkk 1 - \frac{1}{2} \lmk {\mathcal T} - \tau_n \rmk \lmk 3 \frac{a'({\mathcal T})}{a({\mathcal T})} + \frac{\rho_0'({\mathcal T})}{\rho_0({\mathcal T})} \rmk \rkk \,,
\\
&\simeq \frac{1}{9} \lkk 1 + \lmk {\mathcal T} - \tau_n \rmk \tilde{\beta} c_{\rm NL2} \rkk \,,
\eeq
where we define 
\beq
 &c_{\rm NL2} \equiv - \frac{3}{2\tilde{\beta}} \frac{a'(\tau_*)}{a(\tau_*)} - \frac{1}{2\tilde{\beta}} \frac{\rho_0'(\tau_*)}{\rho_0(\tau_*)} \,.
 \label{eq:CLNs}
\eeq
Here and hereafter, we take $\kappa$ to be constant and absorb its time dependence into $\rho_0$.
These expansions are applicable for any functional form of  $\tilde{\Gamma}(\tau)$.

To further simplify \eqs{eq:Deltas} and (\ref{eq:Deltad}), 
we need to specify the form of $\tilde{\Gamma}(\tau)$.
In Sec.~\ref{sec:analytic-delta}, we specifically consider the case where $\tilde{\Gamma}(\tau)$ has a delta-functional dependence,
and calculate $\Delta^{(s)}$ up to the next-to-leading order terms in the limit $\mathcal{H}_*/\tilde{\beta} \ll 1$.
In Sec.\ref{sec:analytic-exp},
we instead consider the case where $\tilde{\Gamma}(\tau)$ exhibits exponential dependence,
and compute $\Delta^{(s)}$ and $\Delta^{(d)}$ up to the next-to-leading order terms in the same limit. 
We also derive their asymptotic expressions for both large and small values of $k$.

\subsection{Case with delta-function nucleation rate}
\label{sec:analytic-delta}

We first consider the case where the nucleation rate is given by
\beq
\tilde{\Gamma}(\tau) = 
n_{\rm nuc} \delta(\tau - \tau_{\rm nuc}) \,,
\label{eq:Gamma2}
\eeq
where $n_{\rm nuc}$ represents the comoving number density of nucleated bubbles, and $\tau_{\rm nuc}$ denotes the time of bubble nucleation.%
\footnote{
Concrete realizations of this type of transition can be found in Refs.~\cite{Jinno:2023vnr,Zhong:2025xwm}.
}
In this case, 
the function $I_1(\tau)$ defined in \eq{eq:I} becomes $4 \pi (\tau - \tau_{\rm nuc})^3 n_{\rm nuc} / 3$, and hence
the time of the bubble collision (defined by $I_1(\tau_*) = 1$) and the inverse duration of the phase transition (defined by $dI_1/d\tau (\tau_*)$) are given by 
\beq
&\tau_* = \tau_{\rm nuc} + \lmk \frac{3}{4 \pi n_{\rm nuc} } \rmk^{1/3} = \tau_{\rm nuc} + \frac{3}{\tilde{\beta}} \,,
\label{eq:delta_taustar}
\\
&\tilde{\beta} = \lmk 36 \pi n_{\rm nuc} \rmk^{1/3}  \,.
\label{eq:delta_betatilde}
\eeq

The integral over $\tau_n$ in \eq{eq:Deltas} can be evaluated using the delta function, 
which contributes only when $\tau_{\rm nuc} < \tau_{xy}$.
Also, the exponent of the probability, $I_2({\mathcal T},\tau_d,r)$, given by \eq{eq:I2xy}, is computed as 
\beq
 I_2 = I_{\rm delta} ({\mathcal T}, \tau_d, r) = n_{\rm nuc} \frac{\pi}{12 r} \lmk r + 2 ({\mathcal T} - \tau_{\rm nuc}) \rmk^2 \lmk 3 \tau_d^2 - r^2 + 4 r ({\mathcal T} - \tau_{\rm nuc}) \rmk \,,
 \label{eq:Idelta}
\eeq
for $\tau_{\rm nuc} < \tau_{xy}$ 
without relying on any approximation.
Noting that $\tau_{xy} = \mathcal{T} - r/2$ and $\abs{\tau_d} \le r$,
the condition $\tau_{\rm nuc} < \tau_{xy}$ implies $\tau_{\rm nuc} < \mathcal{T}$ and $\abs{\tau_d} < r < 2 \mathcal{T} - 2 \tau_{\rm nuc}$. 
The integrals over $\mathcal{T}$, $\tau_d$, and $r$ in \eq{eq:integrationdomain} is therefore restricted to these domains.

We note that the difference between $({\mathcal T}- \tau_{\rm nuc})$ and $({\mathcal T} - \tau_*)$ in the integrand plays a crucial role, since the integral is dominated in the vicinity of ${\mathcal T} \approx \tau_{\rm nuc}$.
We then obtain
\beq
\Delta^{(s)} 
(k/\tilde{\beta}) &\simeq 6 \tilde{\beta}^2 k^3
\int_{\tau_{\rm nuc}}^\infty d {\mathcal T}
\int_0^{2{\mathcal T} - 2 \tau_{\rm nuc}} d\tau_d \; 
\cos(k \tau_d)
 \int_{\tau_d}^{2{\mathcal T} - 2 \tau_{\rm nuc}} d r \, r \, e^{ -I_{\rm delta} ({\mathcal T}, \tau_d, r)}
\nonumber\\
&
 \times \frac{\tilde{\Gamma}_0 }{9} 
\left[ j_0 (k r){\mathcal S}_0 + \frac{j_1 (k r)}{k r}{\mathcal S}_1 + \frac{j_2 (k r)}{k^2r^2}{\mathcal S}_2 \right]_{ \tau_n = \tau_{\rm nuc}} 
\lkk 1 + 3 c_{\rm NL1} + (c_{\rm NL2}-c_{\rm NL1}) \tilde{\beta} ({\mathcal T} - \tau_{\rm nuc}) \rkk \,,
\label{eq:Deltas-flat-delta}
\eeq
for the terms up to next-to-leading order in the expansion around the Minkovski limit,
where $I_{\rm delta} ({\mathcal T}, \tau_d, r)$ is given by \eq{eq:Idelta}. 
Here, we use $\tau_* = \tau_{\rm nuc} + 3 /\tilde{\beta} \simeq \tau_{\rm nuc}$ at leading order. 
Equation~\eqref{eq:Deltas-flat-delta} gives the expression for the case of \eqref{eq:Gamma2} in the limit $\tilde{\beta}/\mathcal{H}_* \gg 1$.

If we further take the large $k$ limit with ${\mathcal T} - \tau_{\rm nuc} \sim 1/\tilde{\beta} \gg 1/k$ and following a similar argument in Sec.~\ref{sec:largek}, we obtain
\beq
\Delta^{(s)} 
(k/\tilde{\beta} \gg 1) &\simeq \frac{4 \pi \tilde{\beta}^2 n_{\rm nuc}^2 }{9 k} 
\int_{\tau_{\rm nuc}}^\infty d {\mathcal T}
 e^{-I_{\rm delta} ({\mathcal T}, 0, 0)}
 ({\mathcal T} - \tau_{\rm nuc})^6 
 \lkk 1 + 3 c_{\rm NL1} + (c_{\rm NL2}-c_{\rm NL1}) \tilde{\beta} ({\mathcal T} - \tau_{\rm nuc}) \rkk
\\
&= \frac{\Gamma(7/3) \tilde{\beta} }{4 \pi k} 
\lmk 1 + 3 c_{\rm NL1} + \frac{3 \, \Gamma(8/3)}{ \Gamma(7/3)} (c_{\rm NL2}-c_{\rm NL1}) \rmk 
\\
&\simeq 0.09475 \frac{ \tilde{\beta} }{k} \lmk 1 + 3 c_{\rm NL1} + 3.791 (c_{\rm NL2}-c_{\rm NL1}) \rmk  \,,
\label{eq:delta-flat-largek}
\eeq
where $\Gamma(x)$ is the Gamma function, and
we use $\tilde{\beta} = (36 \pi n_{\rm nuc})^{1/3}$ in the second line.
On the other hand,
in the small $k$ limit and following a similar argument in Sec.~\ref{sec:smallk}, we obtain
\beq
\Delta^{(s)} 
(k/\tilde{\beta} \ll 1) &\simeq 0.1406 \frac{k^3}{\tilde{\beta}^3} \lmk 1 + 3 c_{\rm NL1} + 3.833 (c_{\rm NL2}-c_{\rm NL1}) \rmk \,.
\label{eq:delta-flat-smallk}
\eeq

As a reference, in the radiation-dominated era with constant $\rho_0$, we obtain
$3 c_{\rm NL1} + 3.791 (c_{\rm NL2}-c_{\rm NL1}) \simeq -0.94 (\mathcal{H}_*/\tilde{\beta})$ and $3 c_{\rm NL1} + 3.833 (c_{\rm NL2}-c_{\rm NL1}) \simeq -0.75 (\mathcal{H}_*/\tilde{\beta})$.
This indicates that the next-to-leading-order terms introduce corrections of about $10\%$ when $\tilde{\beta} / \mathcal{H}* = \mathcal{O}(10)$.
We also note that $3 c_{\rm NL1} = - 18 (\mathcal{H}_*/\tilde{\beta})$ and $3.8 (c_{\rm NL2}-c_{\rm NL1}) \simeq 17 (\mathcal{H}_*/\tilde{\beta})$, which implies that the expansion around the Minkovski limit breaks down for $\tilde{\beta} / \mathcal{H}* \lesssim 20$, unless there are significant cancellations between these terms. Such a situation is expected, for example, in the case of an exponential nucleation rate, as discussed in the next subsection.

\subsection{Case with exponential nucleation rate}
\label{sec:analytic-exp}

We now consider 
the case of an exponential nucleation rate: 
\beq
 \tilde{\Gamma}(\tau) 
 &= \tilde{\Gamma}_0 e^{\tilde{\beta}' \tau} \,,
 \label{eq:gammatilde-exp0}
\eeq
where $\tilde{\Gamma}_0$ and $\tilde{\beta}'$ are constants. 
Note that the parameter $\tilde{\beta}'$ coincides with the one defined in \eq{eq:beta} in the limit where $\tilde{\beta}/\mathcal{H}_* \gg 1$. In particular, we have
\beq
 \tilde{\beta}/\mathcal{H}_* = \tilde{\beta} \tau_* \simeq \log \lkk \frac{ \tilde{\beta}'^4}{8 \pi \tilde{\Gamma}_0} \rkk 
 \qquad {\rm for} \ \tilde{\beta}/\mathcal{H}_* \gg 1 \,.
 \label{eq:approxbetatau}
\eeq

In the limit of large $\tilde{\beta}\tau_*$, we can approximately extend the lower bound of the $\tau$ integral in \eq{eq:I2xy} from $0$ to $-\infty$ because $\tilde{\Gamma}$ is exponentially suppressed in the extended domain. 
This leads to the following expression:
\beq
 &I_2(\tau_x,\tau_y,r) \simeq \frac{8\pi \tilde{\Gamma}({\mathcal T})}{\tilde{\beta}^4} \mathcal{I}(\tau_d, r),
 \label{eq:Ixy}
 \\
 &\mathcal{I}(\tau_d, r) =\lkk \lmk e^{\tilde{\beta} \tau_d/2} + e^{-\tilde{\beta} \tau_d/2} \rmk - \frac{1}{4r} \lmk 4r +r^2\tilde{\beta} - \tau_d^2\tilde{\beta} \rmk e^{-\tilde{\beta} r/2} \rkk \,,
\eeq
for large $\tilde{\beta}/\mathcal{H}_*$. 
In this limit, the dependence on ${\mathcal T}$ factorizes.

Combining them, we obtain 
\beq
\Delta^{(s)} 
(k/\tilde{\beta}) &\simeq 6 \tilde{\beta}^2 k^3
\int_0^\infty d {\mathcal T}
\int_0^{2 {\mathcal T}} d\tau_d \; 
\cos(k \tau_d) 
 \int_{\tau_d}^{2 {\mathcal T}} d r \, r 
\, e^{-\frac{8\pi \tilde{\Gamma}({\mathcal T})}{\tilde{\beta}^4} \mathcal{I}(\tau_d, r) }
\nonumber\\
&\times\int_{0}^{\tau_{xy}} d\tau_n \tilde{\Gamma}(\tau_n) 
\frac{1}{9} \lkk 1 + c_{\rm NL1} \tilde{\beta} (\tau_* - {\mathcal T}) + c_{\rm NL2} \tilde{\beta} ({\mathcal T} - \tau_n) \rkk 
 \nonumber\\
 &\times 
\left[ j_0 (k r){\mathcal S}_0 + \frac{j_1 (k r)}{k r}{\mathcal S}_1 + \frac{j_2 (k r)}{k^2r^2}{\mathcal S}_2 \right] \,,
\eeq
for the single bubble contribution. 
The $\tau_n$ integral can be evaluated, yielding 
\beq
&\int_{0}^{\tau_{xy}} d\tau_n \tilde{\Gamma}(\tau_n) 
 \frac{1}{9} \lkk 1 + c_{\rm NL1} \tilde{\beta} (\tau_* - {\mathcal T}) + c_{\rm NL2} \tilde{\beta} ({\mathcal T} - \tau_n) \rkk
\sum_{i=0}^2 \frac{j_i(kr)}{(kr)^i} {\mathcal S}_i
\nonumber\\
&= \frac{1}{9 \tilde{\beta}^9 r^4} \tilde{\Gamma}({\mathcal T}) e^{-\tilde{\beta} r/2}
\sum_{i=0}^2 \frac{j_i(kr)}{(kr)^i} \lmk (1 + c_{\rm NL1} \tilde{\beta} (\tau_* - {\mathcal T})) F_i^{(\rm L)}(\tilde{\beta} \tau_d, \tilde{\beta} r) + c_{\rm NL2} F_i^{(\rm NL)} (\tilde{\beta} \tau_d, \tilde{\beta} r) \rmk \,,
\label{eq:tauintegral1}
\eeq
where we again extend the lower bound of the integral to $-\infty$ because $\tilde{\Gamma}$ is exponentially suppressed in the extended domain. 
Here, the leading-order terms $F_i^{(\rm L)}$ are given by~\cite{Jinno:2016vai} 
\beq
 &F_0^{(\rm L)} (\tau_d, r) = 2(r^2 - \tau_d^2)^2 (r^2 + 6 r + 12) \,,
 \\
 &F_1^{(\rm L)} (\tau_d, r) = 2 (r^2 - \tau_d^2) \lkk - r^2 ( r^3 + 4 r^2 + 12 r + 24) 
 \right. \nonumber\\
 &\qquad \left.
 + \tau_d^2 ( r^3 + 12 r^2 + 60 r + 120 ) \rkk \,,
 \\
 &F_2^{(\rm L)} (\tau_d, r) = \frac12 \lkk r^4 ( r^4 + 4 r^3 + 20 r^2 + 72 r + 144) 
\right. \nonumber\\
 &\qquad \left.
 - 2 \tau_d^2 r^2 ( r^4 + 12 r^3 + 84 r^2 + 360 r + 720 ) 
 \right. \nonumber\\
 &\qquad \left.
 + \tau_d^4 ( r^4 + 20 r^3 +180 r^2 + 840 r + 1680 ) \rkk, 
\eeq
and 
the next-to-leading order temrs are given by 
\beq
 &F_0^{(\rm NL)}(\tau_d, r) =
 \lmk r^2 -\tau_d^2 \rmk^2 \lmk r^3 + 12 r^2 
 + 60 r +120 \rmk \,,
 \\
 &F_1^{(\rm NL)}(\tau_d, r) = 
 \lmk r^2 -\tau_d^2 \rmk 
 \lkk 
 \tau_d^2 \lmk r^4 + 16 r^3 + 132 r^2 + 600 r + 1200 \rmk
 \right. \nonumber\\
 &\qquad \left. - r^2 \lmk r^4 + 8 r^3 + 36 r^2 + 120 r + 240 \rmk \rkk\,,
 \\
 &F_2^{(\rm NL)}(\tau_d, r) = \frac{1}{4} 
 \lkk 
 r^4 \lmk r^5 + 6 r^4 + 36 r^3 + 192 r^2 + 720 r + 1440 \rmk 
 \right.
 \nonumber\\
 &\qquad - 2 r^2 \tau_d^2 \lmk r^5 + 14 r^4 + 132 r^3 + 864 r^2 + 3600 r + 7200 \rmk 
 \nonumber\\
 &\qquad \left.+ \tau_d^4 \lmk r^5 + 22 r^4 + 260 r^3 + 1920 r^2 + 8400 r + 16800 \rmk
 \rkk\,.
\eeq

Then we note that the $\mathcal{T}$ dependence in the integrand contains a factor
$\exp \lkk -\frac{8\pi \tilde{\Gamma}({\mathcal T})}{\tilde{\beta}^4} \mathcal{I}(\tau_d, r) \rkk$ as well as $\tilde{\Gamma}(\mathcal{T})$. 
The latter favors larger values of $\mathcal{T}$, whereas the former induces a strong suppression once
$\tilde{\beta} \mathcal{T} \gtrsim \ln (\tilde{\beta}^4/(8\pi \tilde{\Gamma}_0 \mathcal{I}(\tau_d,r))) \simeq \tilde{\beta} / \mathcal{H}_* - \ln \mathcal{I}(\tau_d,r)$.
This implies two consequences:
i) the integral over $\mathcal{T}$ is dominated near this lower bound, and therefore $\tilde{\beta} \mathcal{T} \gg 1$ in the limit of large $\tilde{\beta} / \mathcal{H}_*$;
ii) the integral over $\tau_d$ is strongly suppressed when $\mathcal{I}(\tau_d,r) \gg 1$. 
Since $\mathcal{I}(\tau_d,r)$ contains a factor $e^{\tilde{\beta} \tau_d/2}$, this further implies that the upper bound of the $\tau_d$ integration can be extended from $2\mathcal{T}$ to $\infty$, because the additinal region is exponentially suppressed.
Similarly, the upper bound of the $r$ integration can be extended from $2\mathcal{T}$ to $\infty$, because the additional region is exponentially suppressed by the factor $e^{- \tilde{\beta}r/2}$ appearing in \eq{eq:tauintegral1}.

Once we approximately modify the integration domain,
the integral over ${\mathcal T}$ can be performed analytically
by making use of
\beq
 &\int_0^\infty d Y e^{-X e^Y + n Y} \simeq \int_{-\infty}^\infty d Y e^{-X e^Y + n Y} = \frac{(n-1)!}{ X^n}\,,
 \\
 &\int_0^\infty dY e^{-X e^Y + Y} Y \simeq \int_{-\infty}^\infty dY e^{-X e^Y + Y} Y = -\frac{\gamma_E + \log X}{X},
\eeq
with $\gamma_E$ ($\simeq 0.577$) being the Euler's constant, 
where we extend the integration domain into the exponentially suppressed region $Y < 0$.
Recallling \eq{eq:approxbetatau} in a large $\tilde{\beta} \tau_*$ limit, 
we can also approximate 
\beq
 \tilde{\beta} \tau_*  + \log \lmk \frac{8 \pi}{\tilde{\beta}^4} \tilde{\Gamma}_0 \mathcal{I}(\tau_d,r) \rmk 
 \simeq  \log \mathcal{I} \,.
\eeq
We then obtain 
\beq
\Delta^{(s)} 
(k/\tilde{\beta}) &\simeq \frac{k^3}{12\pi \tilde{\beta}^3}
\int_0^\infty d\tau_d 
 \int_{\tau_d}^\infty d r \, r 
 \frac{e^{-\tilde{\beta} r/2} \cos(k \tau_d) }{\tilde{\beta} r^4 \mathcal{I}} 
 \nonumber \\
 &\qquad \times\sum_{i=0}^2 \frac{j_i(kr)}{(kr)^i} \lkk \lmk 1 + c_{\rm NL1} (\gamma_E + \log \mathcal{I}(\tau_d,r)) \rmk F_i^{(\rm L)} (\tilde{\beta} \tau_d, \tilde{\beta} r) 
 \right.
 \nonumber\\
 &\qquad \qquad \qquad \qquad \qquad \qquad\left. + c_{\rm NL2} F_i^{(\rm NL)} (\tilde{\beta} \tau_d, \tilde{\beta} r) \rkk \,,
\label{eq:resultflat}
\eeq
where we use $\tilde{\beta}' \simeq \tilde{\beta}$ at the leading order. 
Equation~(\ref{eq:resultflat}) is an analytic formula for the case with the exponential nucleation rate, including the next-to-leading order correction from the expantion of the Universe. 
The leading order term reproduces the existing formula in Ref.~\cite{Jinno:2016vai}.

For the double bubble contribution, 
we have 
\beq
{\mathcal D}^{(d)}_x(\tau_x,\tau_y, r) &\simeq \pi \kappa \rho_0(\tau_*) \int_{-\infty}^{\tau_{xy}} d\tau_{xn} \tilde{\Gamma} (\tau_{xn}) 
r_x^3 
\lkk 1 + \frac{1}{2} c_{\rm NL2} \tilde{\beta} r_x 
\rkk
\lmk \cos^3 \theta_{x\times} - \cos \theta_{x\times} \rmk, 
\\
&= \frac{\pi \kappa \rho_0(\tau_*)}{6}\frac{e^{-\tilde{\beta} r/2}}{\tilde{\beta}^7 r^3} \Gamma({\mathcal T}) \lmk G_{\rm L}(\tilde{\beta} \tau_d,\tilde{\beta} r) + \frac{1}{2} c_{\rm NL2} G_{\rm NL}(\tilde{\beta} \tau_d,\tilde{\beta} r) \rmk, 
\\
{\mathcal D}^{(d)}_y(\tau_x,\tau_y, r) &\simeq \frac{\pi \kappa \rho_0(\tau_*)}{6}\frac{e^{-\tilde{\beta} r/2}}{\tilde{\beta}^7 r^3} \Gamma({\mathcal T}) \lmk G_{\rm L}(-\tilde{\beta} \tau_d,\tilde{\beta} r) + \frac{1}{2} c_{\rm NL2} G_{\rm NL}(-\tilde{\beta} \tau_d,\tilde{\beta} r) \rmk, 
\eeq
in the Minkovski limit, 
where we again extend the lower bound of the integral to $-\infty$ and define
\beq
 &G_{\rm L}(\tau_d,r) \equiv \lmk r^2 - \tau_d^2 \rmk
 \lkk r^3 +2 r^2 + \tau_d \lmk r^2 + 6 r + 12 \rmk \rkk, 
 \\
 &G_{\rm NL}(\tau_d,r) \equiv 
 \frac{1}{2} \lmk r^2 - \tau_d^2 \rmk
 \lkk r^2 (r + \tau_d)^2 + 6 r (r + \tau_d)^2 + 12 (r^2 + 4 r \tau_d + \tau_d^2) + 96 \tau_d \rkk \,.
\eeq
By applying a similar argument to $\Delta^{(s)}$, 
we finally obtain 
\begin{align}
\Delta^{(d)}(k/\tilde{\beta}) &\simeq 
\frac{k^3}{96 \pi \tilde{\beta}^3}
\int_0^\infty d\tau_d
\int_{\tau_d}^\infty dr \;
\frac{e^{-\tilde{\beta} r} \cos(k\tau_d) }{\tilde{\beta}^2 r^4 \mathcal{I}^2}
 \frac{j_2 (k r)}{k^2r^2} 
 \nonumber\\
 &\times \lmk G_{\rm L}(\tilde{\beta} \tau_d,\tilde{\beta} r) + \frac{1}{2} c_{\rm NL2} G_{\rm NL}(\tilde{\beta} \tau_d,\tilde{\beta} r) \rmk
 \lmk G_{\rm L}(-\tilde{\beta} \tau_d,\tilde{\beta} r) + \frac{1}{2} c_{\rm NL2} G_{\rm NL}(-\tilde{\beta} \tau_d,\tilde{\beta} r) \rmk
 \,.
\label{eq:Deltaflat}
\end{align}
\eqs{eq:resultflat} and (\ref{eq:Deltaflat}) 
provide the expressions for the nucleation rate of \eq{eq:gammatilde-exp0} in the limit $\tilde{\beta}/\mathcal{H}_* \gg 1$.

Furthermore, in the large-$k$ limit,
all the integrals in \eq{eq:resultflat} and (\ref{eq:Deltaflat}) can be evaluated analytically or numerically, yielding
\beq
 &\Delta^{(s)} 
(k/\tilde{\beta} \gg 1) \simeq \frac{\tilde{\beta}}{3 \pi k} \lmk 1 + \gamma_E c_{\rm NL1} + 5 c_{\rm NL2}  \rmk, 
\label{eq:deltas-exp-largek}
\\
 &\Delta^{(d)} 
(k/\tilde{\beta} \gg 1) \simeq \frac{4 \tilde{\beta}^2}{35 k^2} \lmk 1 + 4 c_{\rm NL2} \rmk. 
\eeq
The leading terms agree with those in Ref.\cite{Jinno:2016vai} within a precision of approximately 5\%.
In the small-$k$ limit, 
we obtain 
\beq
&\Delta^{(s)} 
(k/\tilde{\beta} \ll 1) 
\simeq 0.2792 \frac{k^3}{\tilde{\beta}^3} \lmk 1 + 0.2608 c_{\rm NL1} + 6.679 c_{\rm NL2} \rmk \,,
\label{eq:deltas-exp-smallk}
\\
&\Delta^{(d)} 
(k/\tilde{\beta} \ll 1) 
\simeq 0.1028 \frac{k^3}{\tilde{\beta}^3} \lmk 1 + 5.636 c_{\rm NL2} \rmk \,.
\eeq
The leading terms agree with those in Ref.~\cite{Jinno:2016vai},
within a precision of approximatrely 10\%,
which falls within the numerical precision reported in that work.
Note that our results can be used for arbitrary functions of $a(\tau)$ and $\rho_0(\tau)$.

As a reference, in the radiation-dominated era with constant $\rho_0$, we obtain $\gamma_E c_{\rm NL1} + 5 c_{\rm NL2} \simeq 11 \mathcal{H}_*/\tilde{\beta}$ 
and $0.2608 c_{\rm NL1}  + 6.679 c_{\rm NL2} \simeq 12 \mathcal{H}_*/\tilde{\beta}$, 
and the next-to-leading order terms introduce corrections of about $10\%$ when $\tilde{\beta}/\mathcal{H}_* \sim 120$. 
This also implies that 
the expansion breaks down for $\tilde{\beta}/\mathcal{H}_* \lesssim 12$ in this case.

\section{Numerical results}
\label{sec:numerical}

In this section, we present the behavior of the GW spectrum evaluated from Eqs.(\ref{eq:Deltas}) and (\ref{eq:Deltad}) without relying on approximations and check the consistency with the results of asymptotic formula.
We specifically consider the case of a radiation-dominated epoch, where $a(\tau) /a_* = \tau /\tau_*$, and assume $\kappa(\tau) \rho_0(\tau) = \kappa \rho_0 = ({\rm const.})$.
In this case, we obtain
\beq
 \rho_B(\tau;\tau_n) = \frac{\kappa \rho_0 (\tau-\tau_n)}{l_B} 
 \lkk \frac{1}{6} + \frac{1}{10} \lmk \frac{\tau_n}{\tau} \rmk + \frac{1}{20} \lmk \frac{\tau_n}{\tau}\rmk^2 + \frac{1}{60} \lmk \frac{\tau_n}{\tau} \rmk^3 \rkk \,.
\eeq
We consider the delta-function bubble nucleation rate in Sec.\ref{sec:result_delta}, and the exponential nucleation rate in Sec.\ref{sec:result_exp}.

The multidimensional numerical integrations (\ref{eq:Deltas}) and (\ref{eq:Deltad}) were performed using the Vegas algorithm~\cite{Lepage:1977sw} as implemented in the CUBA library~\cite{Hahn:2004fe}.
Vegas is a Monte Carlo algorithm that employs importance sampling to reduce variance.
The calculations were carried out using the Mersenne Twister pseudo-random number generator, with a requested relative accuracy of $3 \times 10^{-3}$.

\subsection{Case with delta-function nucleation rate}
\label{sec:result_delta}

\begin{figure}
 \centering
 \includegraphics[width=0.75\hsize]{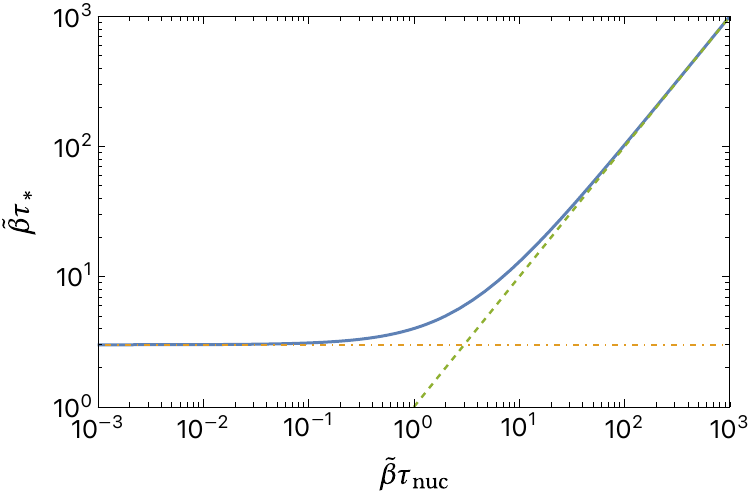}
 \caption{
 $\tilde{\beta} \tau_*$ (blue solid curve) as a function of $\tilde{\beta} \tau_{\rm nuc}$ for the case of a delta-function nucleation rate. 
 The dashed and dot-dashed lines represent the asymptotic forms for large and small values of $\tilde{\beta} \tau_{\rm nuc}$, respectively. 
 Note that $\tau_* = \mathcal{H}_*$. 
}
 \label{fig:betaH}
\end{figure}

First, we consider the delta-functional nucleation rate 
given by \eq{eq:Gamma2}: 
\beq
 \tilde{\Gamma}(\tau) = n_{\rm nuc} \, \delta(\tau - \tau_{\rm nuc}) \,.
\eeq
In this case, the time of the bubble collision and the inverse duration of the phase transition are given by Eqs.~(\ref{eq:delta_taustar}) and (\ref{eq:delta_betatilde}). 
Figure~\ref{fig:betaH} shows $\tilde{\beta} \tau_*$ (blue solid curve) as a function of $\tilde{\beta} \tau_{\rm nuc}$. The dashed line indicates the asymptotic behavior in the limit of large $\tilde{\beta} \tau_{\rm nuc}$, given by $\tilde{\beta} \tau_* = \tilde{\beta} \tau_{\rm nuc}$, while the dot-dashed line corresponds to the asymptotic behavior in the limit of small $\tilde{\beta} \tau_{\rm nuc}$, given by $\tilde{\beta} \tau_* = 3$.

Before presenting the results of numerical calculations, we discuss the parameter dependence of the GW spectrum. The theory contains two free parameters: $n_{\rm nuc}$ and $\tau_{\rm nuc}$. Consequently, the GW spectrum depends on $k$, $n_{\rm nuc}$, and $\tau_{\rm nuc}$. Using Eqs.~(\ref{eq:delta_taustar}) and (\ref{eq:delta_betatilde}), the parameters $n_{\rm nuc}$ and $\tau_{\rm nuc}$ can be rewritten in terms of $\tilde{\beta}$ and $\tau_*$. After rescaling the dimensionful parameters, we find that a single parameter can be trivially factored out, and the quantity $\Delta^{(i)}$ depends only on the combinations $k/\tilde{\beta}$ and $\tilde{\beta} \tau_*$.

\begin{figure}
\centering
\includegraphics[width=0.75\hsize]{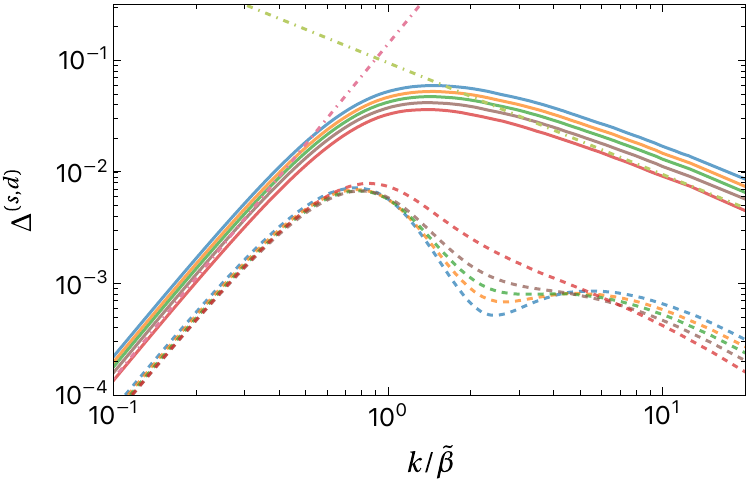}
\caption{
GW spectra $\Delta^{(s)}$ (solid curves) and $\Delta^{(d)}$ (dashed curves) as functions of $k/\tilde{\beta}$ for $\tilde{\beta} \tau_{\rm nuc} = 0.1$ (blue), $0.5$ (orange), $1$ (green), $2$ (brown), and $10$ (red), in the case of delta-function nucleation rate. 
The dot-dashed lines represent the asymptotic behavior corresponding to the leading-order result in the Minkovski limit, as derived in Eqs.~(\ref{eq:delta-flat-largek}) and (\ref{eq:delta-flat-smallk}).
}
\label{fig:delta1}
\end{figure}

For the purpose of illustration, we treat $\tilde{\beta} \tau_{\rm nuc}$, rather than $\tilde{\beta} \tau_*$, as a free parameter in this subsection.
Note that $\tilde{\beta} \tau_*$ can be expressed in terms of $\tilde{\beta} \tau_{\rm nuc}$ via \eq{eq:delta_taustar}.
In Fig.\ref{fig:delta1}, we plot the values of $\Delta^{(s)}$ (solid) and $\Delta^{(d)}$ (dashed) as functions of $k/\tilde{\beta}$ for $\tilde{\beta} \tau_{\rm nuc} = 0.1$ (blue), $0.5$ (orange), $1$ (green), $2$ (brown), and $10$ (red).
The yellow and magenta dot-dashed lines indicate the asymptotic behaviors for the leading result in the Minkovski limit, given by Eqs.~(\ref{eq:delta-flat-largek}) and (\ref{eq:delta-flat-smallk}), respectively.
The full numerical results are in agreement with these asymptotic forms in the limits $k/\tilde{\beta} \ll 1$ and $k/\tilde{\beta} \gg 1$.
It is also evident that the double-bubble contribution is significantly smaller than the single-bubble contribution, as in the Minkovski case.

We also find that the shape of the spectrum from a single bubble, $\Delta^{(s)}$, remains nearly identical across all cases, differing mainly in the overall amplitude by a small amount. These results suggest that the GW spectrum depends primarily on the ratio $k/\tilde{\beta}$ and only weakly on $\tilde{\beta} \tau_{\rm nuc}$ (or equivalently, $\tilde{\beta} \tau_*$), as discussed in Ref.~\cite{Yamada:2025hfs}.

\begin{figure}
 \centering
 \includegraphics[width=0.45\hsize]{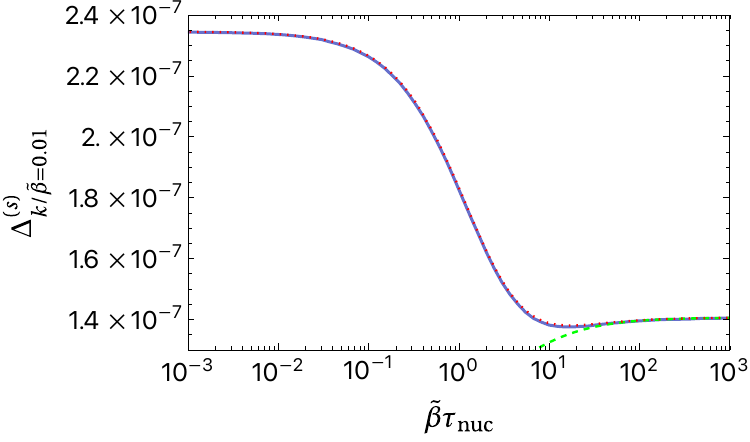}
 \quad 
 \includegraphics[width=0.45\hsize]{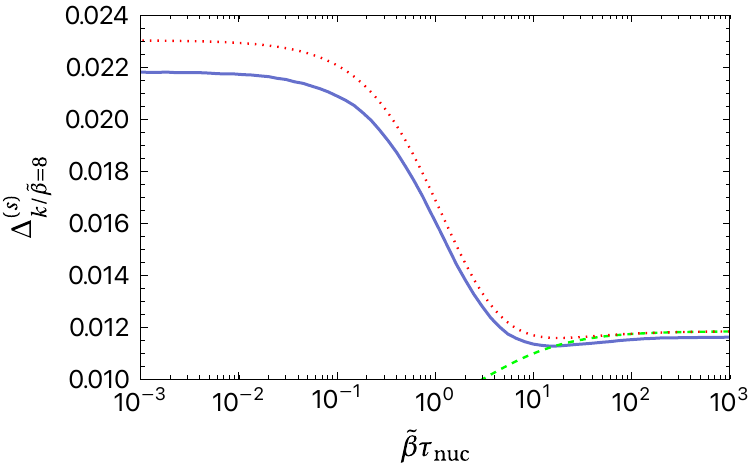}
 \caption{
 Amplitude of the single-bubble contribution $\Delta^{(s)}$ for $k/\tilde{\beta} = 0.01$ ((left panel) and $8.0$ (right panel) as a function of $\tilde{\beta} \tau_{\rm nuc}$ in the case of a delta-function nucleation rate.
The red dotted curves represent the results obtained using the large- and small-$k$ limits discussed in Sec.~\ref{sec:asymptotic}. 
In the left panel, the blue solid curve completely overlaps with the red dotted curve.
The green dashed curves correspond to the analytic next-to-leading order results in the large $\tilde{\beta}/\mathcal{H}_*$ limit, as discussed in Sec.\ref{sec:analytic-delta}.
}
 \label{fig:delta3}
\end{figure}

Figure~\ref{fig:delta3} shows the GW amplitude from the single-bubble contribution at small and large wavenumbvers, illustrating consistency with the asymptotic results discussed in Sec.~\ref{sec:asymptotic} and \ref{sec:largebeta}. We consider two representative cases: $k/\tilde{\beta} = 0.01$ (left panel) and $8.0$ (right panel), corresponding to small and (moderately) large wavenumbers, respectively.
The solid blue curve represents the full numerical result, while the red dotted curve shows the result obtained using the small (left panel) or large (right panel) wavenumber limits described in Sec.~\ref{sec:asymptotic}. In the left panel, the two curves are nearly indistinguishable, indicating excellent agreement at sufficiently small wavenumber relative to $\tilde{\beta}$. In the right panel, although $k/\tilde{\beta}$ is not extremely large, the results agree within approximately $5\%$. Although the agreement improves further at larger wavenumbers, the computational cost of full numerical calculations increases significantly in that regime.

The green dashed curve in Fig.~\ref{fig:delta3} shows the result obtained using the expansion around the Minkovski limit up to next-to-leading-order terms, combined with the large- or small-wavenumber limits. The corresponding formulas are given in \eq{eq:delta-flat-largek} and \eq{eq:delta-flat-smallk}. This figure demonstrates that the asymptotic formulas in the Minkovski limit reproduce the full numerical results to within about $5\%$ accuracy for $\tilde{\beta} \tau_{\rm nuc} \gtrsim 10$.

Although the inclusion of the next-to-leading-order term leads to larger deviations at $\tilde{\beta} \tau_{\rm nuc} \lesssim 10$ compared with the leading-order case, it provides a better fit for $\tilde{\beta} \tau_{\rm nuc} \gtrsim 30$. This suggests that next-to-next-to-leading-order terms become relevant around $\tilde{\beta} \tau_{\rm nuc} = \mathcal{O}(10)$ and that the expansion breaks down below this regime. A likely explanation is a nontrivial cancellation among different contributions within the next-to-leading-order terms for this specific case, as discussed at the end of Sec.~\ref{sec:analytic-delta}.

\subsection{Case with exponential nucleation rate}
\label{sec:result_exp}

Next, we consider 
the nucleation rate given by \eq{eq:gammatilde-exp0}: 
\beq
 \tilde{\Gamma}(\tau) 
 &= \tilde{\Gamma}_0 e^{\tilde{\beta}' \tau} \,.
 \label{eq:nucleation-rate-exp}
\eeq
Even though the nucleation rate diverges in the limit $\tau \to \infty$,
we do not need to introduce an ad hoc cutoff for the $\tau$ integral in \eq{eq:Deltas}.
As the transition approaches completion, $P_2$ decays exponentially to zero, effectively shutting off the GW source.
In contrast,
we should note that the exponential nucleation rate does not vanish in the limit $\tau \to 0$. 
In cases where $\tilde{\beta} \tau_*$ is relatively small, bubble nucleation may predominantly occur before the exponential dependence becomes significant.
This can be examined by computing $\tau_*$ from the condition $I_1(\tau_*) = 1$ using \eq{eq:I}, and checking whether the exponential dependence dominates the integral.
We find that the exponential behavior becomes dominant only if $\tilde{\beta} \tau_* \gtrsim 10$.

\begin{figure}
 \centering
 \includegraphics[width=0.75\hsize]{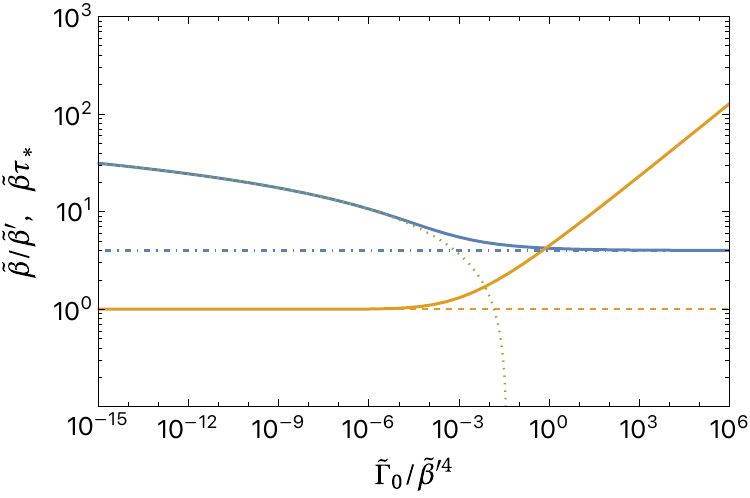}
 \caption{
 $\tilde{\beta}/\tilde{\beta}'$ (orange curve) and $\tilde{\beta} \tau_*$ (blue curve) as functions of $\tilde{\Gamma}_0 / \tilde{\beta}'^4$. 
 The dashed and dot-dashed lines indicate the asymptotic values, while the dotted curve represents the approximation given in \eq{eq:approxbetatau}.
}
 \label{fig:betatau}
\end{figure}

Figure~\ref{fig:betatau} shows the behavior of $\tilde{\beta}/\tilde{\beta}'$ (orange curve) and $\tilde{\beta} \tau_*$ (blue curve) as functions of $\tilde{\Gamma}_0 / \tilde{\beta}'^4$. 
The dashed line corresponds to $\tilde{\beta} = \tilde{\beta}'$. 
As indicated by the dot-dashed line, 
we observe that $\tilde{\beta} \mathcal{H}_*$ ($= \tilde{\beta} \tau_*$) asymptotically approaches 4 in the limit of large $\tilde{\Gamma}_0/ \tilde{\beta}'^4$. 
The dotted curve represents the approximation given in \eq{eq:approxbetatau}.

In addition to the wavenumber $k$, 
there are two free parameters in this case: $\tilde{\Gamma}_0$ and $\tilde{\beta}'$. However, either parameter can be eliminated by rescaling all dimensionful quantities according to their mass dimensions with respect to $\tilde{\beta}'$ (e.g., $k \to \tilde{\beta}' k$, $r \to r/\tilde{\beta}'$, ${\mathcal T} \to {\mathcal T}/\tilde{\beta}'$, and $\tilde{\Gamma}_0 \to \tilde{\beta}'^4 \tilde{\Gamma}_0$.) 
Furthermore, the free parameters can be expressed in terms of $\tilde{\beta}$ and $\tau_*$, as defined by Eqs.~(\ref{eq:I}) and (\ref{eq:beta}). 
These observations imply that the resulting GW spectrum depends only on the combinations $k/\tilde{\beta}$ and $\tilde{\beta} \tau_*$. 
For the purpose of numerical evaluation, we treat $\tilde{\Gamma}_0 / \tilde{\beta}'^4$ as a free parameter (instead of $\tilde{\beta} \tau_*$) and present the results in terms of $k/\tilde{\beta}$.

\begin{figure}
 \centering
 \includegraphics[width=0.75\hsize]{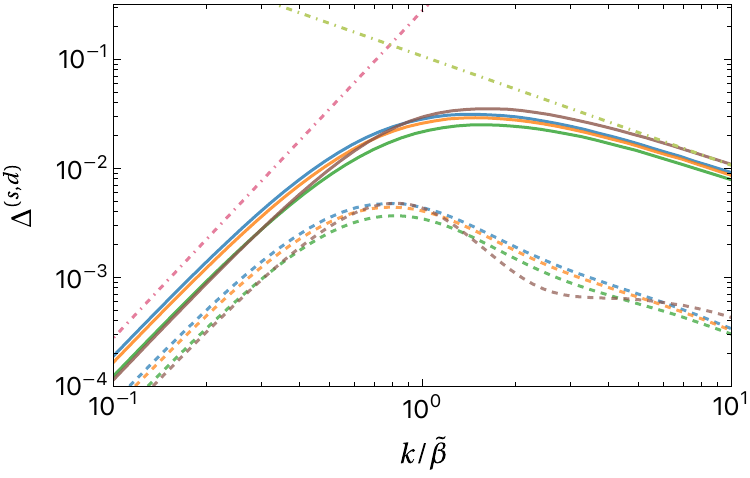}
 \caption{
 Same as Fig.~\ref{fig:delta1} but for the case with an exponential bubble nucleation rate. 
 We take $\tilde{\Gamma}_0/\tilde{\beta}'^4 = 10^{-15}$ (blue), $10^{-11}$ (orange), $10^{-7}$ (green), and $10^1$ (brown). The dot-dashed lines show the asymptotic behavior of the leading-order result in the Minkovski limit, as derived in Eqs.~(\ref{eq:deltas-exp-largek}) and (\ref{eq:deltas-exp-smallk}). 
}
 \label{fig:exp1}
\end{figure}

Figure~\ref{fig:exp1} shows the spectra from the single-bubble contribution (solid) and the double-bubble contribution (dashed).
We consider values of $\tilde{\Gamma}_0/\tilde{\beta}'^4$ equal to $10^{-15}$ (blue), $10^{-11}$ (orange), $10^{-7}$ (green), and $10^1$ (brown).
Note that 
the parameter $\tilde{\Gamma}_0/\tilde{\beta}'^4$ can be expressed in terms of $\tilde{\beta}\tau_*$ ($=\tilde{\beta}/\mathcal{H}_*$), as shown in Fig.~\ref{fig:betatau}, and we find 
$\tilde{\beta}\tau_* = 31$, $22$, $13$, and $4.1$, respectively.

The dot-dashed lines indicate the analytic results for the asymptotic behavior in the Minkovski limit, as derived in Eqs.~(\ref{eq:deltas-exp-largek}) and (\ref{eq:deltas-exp-smallk}). The agreement is not exact even for $\tilde{\Gamma}_0/\tilde{\beta}'^4 = 10^{-15}$, corresponding to $\tilde{\beta}\tau* = 31$. This is consistent with the fact that the next-to-leading-order term contributes a factor of $\simeq 11 /\tilde{\beta}\tau_* \sim 0.4$, indicating that the large $\tilde{\beta} / \mathcal{H}_*$ expansion already breaks down in these examples.

Nevertheless, we emphasize that all results exhibit similar amplitudes, including the case with $\tilde{\beta} / \mathcal{H}_* \simeq 4.1$. This universality, up to an $\mathcal{O}(1)$ factor, is discussed in Ref.~\cite{Yamada:2025hfs}.

\section{Implications for GW observations}
\label{sec:implication}

For readers' convenience, here we briefly explain the relation between physical parameters and conformal parameters utilized throughout this paper.

\begin{figure}
 \centering
 \includegraphics[width=0.75\hsize]{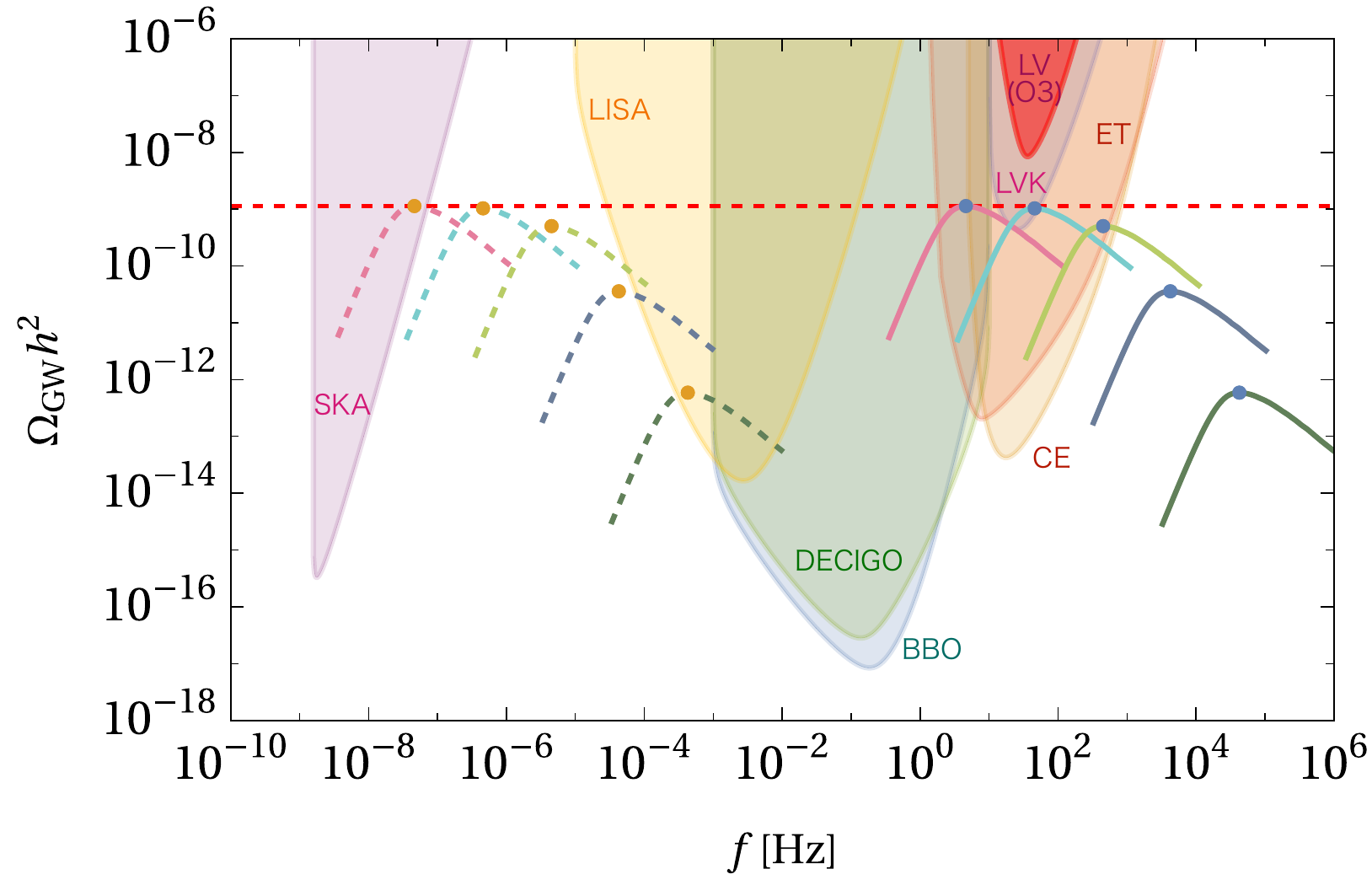}
 \caption{
 GW spectra for the cases with $\beta / H_* -3 = 0.01$, $0.1$, $1$, $10$, and $10^2$ (from left to right). 
 We take $H_{\rm nuc} = 10^{-13} \GeV$ (dashed) and $10^{3} \GeV$ (solid), assuming a delta-function nucleation rate during the radiation-dominated epoch. We set $\alpha_* = 0.1$. The horizontal dashed line denotes the upper bound on the peak amplitude.
 The lightly shaded regions indicate projected sensitivities. The red-shaded region is excluded by the LV O3 data~\cite{KAGRA:2021kbb,LIGOScientific:2021nrg}.
}
 \label{fig:spectrum_peak}
\end{figure}

Specifically, we consider the scenario with a delta-function nucleation rate in the radiation-dominated epoch. The nucleation rate per unit physical volume and time is given by:
\beq
 \Gamma (t) &= \frac{n_{\rm nuc}}{a^3(t_{\rm nuc})} 
 \delta (t - t_{\rm nuc})
 \\
 &\equiv \Gamma_0 \delta (t - t_{\rm nuc})\,,
\eeq
where $\Gamma_0$ is defined by the second line.
In the radiation-dominated epoch, the nucleation time is given by
\beq
 t_{\rm nuc} = \frac{1}{2 H_{\rm nuc}} \,.
\eeq
The Hubble parameter relates to the radiation temperature as 
\beq
 T_* \simeq 100 \GeV \lmk \frac{H_*}{6.9 \times 10^{-14} \GeV} \rmk^{1/2} \lmk \frac{g_*}{100} \rmk^{-1/4}\,,
\eeq
with $g_*$ being the effective relativistic degrees of freedom, 
\beq
 H_* = \lmk \frac{(36 \pi \Gamma_0)^{1/3}}{(36 \pi \Gamma_0)^{1/3} + 3 H_{\rm nuc}} \rmk^2 H_{\rm nuc} \,,
\eeq
and
\beq
 \frac{a(t_{\rm nuc})}{a(t_*)} = \frac{(36 \pi \Gamma_0)^{1/3}}{(36 \pi \Gamma_0)^{1/3} + 3 H_{\rm nuc}}\,.
\eeq
Using Eqs.~(\ref{eq:delta_taustar}) and (\ref{eq:delta_betatilde}),
the ratio $\beta / H_*$ ($=\tilde{\beta} / \mathcal{H}_* = \tilde{\beta} \tau_*$) 
can be expressed in terms of $\Gamma_0$ and $H_{\rm nuc}$ as
\beq
 \frac{\beta}{H_*} = \frac{(36 \pi \Gamma_0)^{1/3}}{H_{\rm nuc}} + 3 \,.
\eeq
These quantities are useful to calculate the GW spectrum via Eqs.~(\ref{eq:f_present}) and (\ref{eq:Omega_present}).

Figure~\ref{fig:spectrum_peak} shows the GW spectrum for the cases with $\beta / H_* -3 = 0.01$, $0.1$, $1$, $10$, and $10^2$ (from left to right). We adopt a representative value of $\alpha_* = 0.1$, though dependence on $\alpha_*$ is straightforward. We take $H_{\rm nuc} = 10^{-13} \GeV$ (dashed) and $10^{3} \GeV$ (solid), corresponding to nucleation temperatures $T (t_{\rm nuc}) \sim 100 \GeV$ and $10^{10} \GeV$, respectively. 
As $\Gamma_0$ decreases, $\beta/H_*$ approaches its asymptotic value of 3, at which the GW amplitude saturates, as shown by the horizontal red dashed line.

If the nucleation temperature and the duration of the phase transition are properly interpreted regarding the parameters for the nucleation rate, results for an exponential nucleation rate are qualitatively similar. However, in such cases, the peak amplitude should be scaled by approximately a factor of $1/3\,\text{-}\,1/2$, by compareing Figs.~\ref{fig:delta1} and \ref{fig:exp1}.

We include projected sensitivity curves from future GW experiments. 
Power-law-integrated sensitivity curves for forthcoming experiments are indicated by lighter shaded regions, following Ref.~\cite{Schmitz:2020syl}. These upcoming experiments include SKA~\cite{Janssen:2014dka}, LISA~\cite{LISA:2017pwj}, DECIGO~\cite{Kawamura:2011zz,Kawamura:2020pcg}, BBO~\cite{Harry:2006fi}, Einstein Telescope (ET)~\cite{Punturo:2010zz,Maggiore:2019uih}, Cosmic Explorer (CE)~\cite{Reitze:2019iox}, and aLIGO+aVirgo+KAGRA (LVK)~\cite{Harry:2010zz,LIGOScientific:2014pky,Somiya:2011np,KAGRA:2020cvd}.
The densely shaded red region is excluded by data from the advanced LIGO/Virgo third observational run (LV O3)~\cite{KAGRA:2021kbb,LIGOScientific:2021nrg}. 
For a strongly supercooled phase transition, the GW amplitude can reach the projected sensitivity of future experiments, including LVK.

\section{Discussion and conclusion}
\label{sec:discussion}

We have extended the analytic formula for the GW spectrum generated by bubble collisions during a FOPT, taking into account the expansion rate of the FLRW Universe.
The effects of cosmic expansion appear through several factors, including the change in the physical volume of the false vacuum, the vacuum energy density (or latent heat), and the redshift of the energy density and frequency of both bubble walls and GWs.
The general analytic formula can be derived in parallel with the original work, once the physical quantities are replaced by their conformal counterparts.

The analytic formula allows for expansions in certain limits, such as the large- and small-wavenumber limits and the Minkovski limit, by expanding the integrand in terms of small parameters.
We provide simplified analytic expressions in these asymptotic regimes, which show good agreement with our full nulerical results as well as existing results in the literature.
These formulas for the large- and small-wavenumber limits are useful for estimating the GW spectrum for a general nucleation rate, since they involve non-oscillatory integrals that can be evaluated much more easily than the original oscillatory expressions.

The expansion around the Minkovski limit (or large $\beta / H_*$ expansion) is calculated up to next-to-leading-order terms for both the delta-function and exponential nucleation rates. In the latter case, the next-to-leading-order term yields about a $10\%$ correction for $\beta / H_* \simeq 140$, which is a typical value for an electroweak phase transition if it proceeds as a first-order phase transition. 
If one aims to determine the peak amplitude of the GW spectrum with precision, the correction remains non-negligible even for such a large value of $\beta / H_*$.

In contrast, our numerical results show that, even in the supercooled regime with $\beta / H_* = \mathcal{O}(1)$, the overall amplitude does not change by more than an $\mathcal{O}(1)$ factor, once the $(\beta / H_*)^{-2}$ dependence is factored out under the appropriate definition of $\beta$, as discussed in the companion paper~\cite{Yamada:2025hfs}.

Finally, we comment on the differences from a related work~\cite{Zhong:2021hgo}, where a similar analytic calculation in the FLRW Universe was performed. The differences are threefold:
i) the form of the bubble energy density, $\rho(x)$, was valid only for a very specific time dependence of the false vacuum energy (see their Eq.~(3.4) and our Eq.~(2.14));
ii) the nucleation rate was assumed to take only the exponential form in their Eq.~(3.7), 
and the expansion history was assumed to be that of the radiation-dominated era; and
iii) the definition of $\alpha$, the ratio between vacuum and radiation energy densities, was not specified around their Eq.~(3.26), even though the ratio is time dependent.
Point (iii) is crucial for interpreting their calculation, since the ratio is strongly time dependent. For example, the ratio scales as $\alpha \propto a^4(t)$ for a constant false vacuum energy during the radiation-dominated era. 
Moreover, we note that their definition of $\Delta^{\rm F}$ does not include a factor of $(\beta/H_*)^2$. 
Consequently, if we denote their result as $\Delta^{\rm F}$ ($\propto \Omega_{\rm GW} /\alpha^2$) and ours as $\Delta^{(i)}$ ($\propto \Omega_{\rm GW} (\beta/H_*)^2/\alpha_*^2$), the two differ by a factor of $(\beta/H_*)^2 (\alpha/\alpha_*)^2$. With our definition of $\alpha_*$, $\Delta^{(i)}$ changes only by a factor of a few for any $\beta/H_*$. By contrast, if $\alpha$ is defined at a different time, $\Delta^{\rm F}$ can vary much more significantly due to the factor $(\alpha/\alpha_*)^2$.
A naive expectation is that, in their calculation, $\alpha$ is evaluated at a time differing by the duration of the phase transition, $\beta^{-1}$.%
\footnote{
Note that their $\eta_*$ is not idential to our $\tau_*$. 
} 
This modifies $\alpha$ by a factor of order $(\alpha/ \alpha_*)^2  \propto (a/a(\tau_*))^8 \sim 1 + 8 \mathcal{O}(H_*/\beta)$ for $\beta/H_* \gg 1$. Because of this prefactor, $8 \mathcal{O}(H_*/\beta)$ can still be of order unity even when $H_*/\beta = \mathcal{O}(10)$. We expect that this explains why their $\Delta^{\rm F}$ appears strongly suppressed for $\beta/H_* \lesssim \mathcal{O}(10)$. The same reasoning may also apply to their ratio $\Delta^{\rm F}/\Delta^{\rm M}$ (see Fig.~5 of their paper), where $\Delta^{\rm M}$ is evaluated in the Minkowski background.

\section*{Acknowledgments}
We thank Ryusuke Jinno for fruitful discussions throughout this work. 
This work is supported by JSPS KAKENHI Grant Number 23K13092.

\appendix

\section{Gravitational waves at present}

The tensor perturbation $h_{ij}$ obeys the equation of motion
\begin{align}
\bar{h}''_{ij}(\tau,\vec{k})+\lmk k^2 - \frac{a''}{a} \rmk \bar{h}_{ij} (\tau,\vec{k}) 
&= {8\pi G } a^3(\tau) \Pi_{ij}(\tau,\vec{k}) \,,
\label{eq:hEOM}
\end{align}
where we define $\bar{h}_{ij} \equiv a(\tau) h_{ij}$ and 
$k^2 = \vec{k}^2$. 
The term $a''(\tau)/a(\tau)$ vanishes during the radiation-dominated epoch and is also negligible for modes well inside the Hubble horizon, where $a''/a \sim a^2 H^2 \ll k^2$.

The transverse-traceless part of the anisotropic stress $\Pi_{ij}$ is computed from the energy-momentum tensor as
\begin{align}
a^2 (\tau) \Pi_{ij}(\tau,\vec{k})
&= K_{ij,kl}(\hat{k}) T_{kl}(\tau,\vec{k}) \,,
\label{eq:PiKT}
\end{align}
where $K_{ij,kl}$ is the projection onto the transverse-traceless part: 
\begin{align}
K_{ij,kl}(\hat{k})
&= P_{ik}(\hat{k})P_{jl}(\hat{k}) - \frac{1}{2}P_{ij}(\hat{k})P_{kl}(\hat{k}) \,, 
\label{eq:kk}
\\
P_{ij}(\hat{k})
&\equiv \delta_{ij}-\hat{k}_i\hat{k}_j \,.
\end{align}

The solution to the equation of motion is formally given by 
\begin{align}
\bar{h}_{ij}(t,\vec{k})
&= 8\pi G
\int_0^\tau d\tau' \;
a^3(\tau')
G_k(\tau,\tau') \Pi_{ij}(\tau',\vec{k})
\;\;\;\;\;\; \,,
\end{align}
where $G_k(\tau,\tau') = \sin (k(\tau - \tau'))/k$ is the Green's function. 
When the source becomes negligible for $\tau > \tau_{\rm end}$, 
the solution can be expressed as
\begin{align}
\bar{h}_{ij}(\tau,\vec{k})
&= A_{ij}(\vec{k}) \sin(k(\tau - \tau_{\rm end})) + B_{ij}(\vec{k}) \cos(k(\tau - \tau_{\rm end})) \,,
\;\;\;\;\;\;
{\rm for} \ \ 
\tau \ge \tau_{\rm end} \,,
\label{eq:hsol}
\end{align}
where the coefficients are given by
\begin{align}
A_{ij}(\vec{k})
&= \frac{8\pi G}{k}\int_0^{\tau_{\rm end}} d\tau' \;
a^3(\tau') \cos(k(\tau_{\rm end} - \tau')) \Pi_{ij}(\tau',\vec{k}) \,, \\
B_{ij}(\vec{k})
&= \frac{8\pi G}{k}\int_0^{\tau_{\rm end}} d\tau' \;
a^3(\tau') \sin(k(\tau_{\rm end} - \tau')) \Pi_{ij}(\tau',\vec{k}) \,.
\end{align}

We define the power spectra as 
\begin{align}
&\langle \bar{h}'_{ij}(\tau,\vec{k}) \bar{h}'^{*}_{ij}(\tau,\vec{q}) \rangle
= (2\pi)^3 \delta^{(3)}(\vec{k} - \vec{q}) P_{\bar{h}'}(\tau,k) \,,
\label{eq:hhspectrum}
\\
&\langle \Pi_{ij}(\tau_x,\vec{k})\Pi^*_{ij}(\tau_y,\vec{q}) \rangle
= (2\pi)^3 \delta^{(3)}(\vec{k} - \vec{q})\Pi(\tau_x,\tau_y,k) \,.
\label{eq:PiPi}
\end{align}
By substituting \eq{eq:hsol} into \eq{eq:hhspectrum} and using \eq{eq:PiPi}, we obtain 
\begin{align}
&P_{\bar{h}'}(\tau_{\rm end},k) 
= 32\pi^2G^2
\int_0^{\tau_{\rm end}} d\tau_x
\int_0^{\tau_{\rm end}} d\tau_y \;
a^3(\tau_x) a^3(\tau_y)
\cos(k(\tau_x - \tau_y))\Pi (\tau_x,\tau_y,k) \,,
\label{eq:Pdoth}
\end{align}
where the result is averaged over the oscillation period.

The energy density of stochastic GWs can be expressed as
\begin{align}
\rho_{\rm GW}(t)
&= \frac{1}{8\pi G} \left< \frac{d {h}_{ij}}{dt} (t,\vec{x}) \frac{d {h}_{ij}}{dt} (t,\vec{x}) \right>
\nonumber \\
&\simeq \frac{1}{8\pi G a^4 (\tau)} \left< \bar{h}'_{ij} (\tau,\vec{x}) \bar{h}'_{ij} (\tau,\vec{x}) \right>
\nonumber \\
&= \frac{1}{8\pi G a^4 (\tau)} \int \frac{d^3 k}{(2\pi)^3} \, P_{\bar{h}'}(\tau,k) \,,
\end{align}
where we have used the condition $k/a \gg H$ for sub-Hubble modes in the second line.
Here, the angle brackets $\left< \cdots \right>$ denote an ensemble average, typically interpreted as a spatial or statistical average over the stochastic background of GWs.
The spectrum of the density parameter for GWs is then given by 
\begin{align}
\Omega_{\rm GW}(\tau,k) &\equiv \frac{1}{\rho_{\rm tot}(\tau)} \frac{d\rho_{\rm GW}}{d\ln k}
\nonumber
\\
&= \frac{2Gk^3}{\pi a^4(\tau) \rho_{\rm tot}(\tau)}
\int_0^{\tau_{\rm end}} d\tau_x
\int_0^{\tau_{\rm end}} d\tau_y \;
a^3(\tau_x) a^3(\tau_y) \cos(k(\tau_x - \tau_y))\Pi (\tau_x,\tau_y,k) \,.
\end{align}

It is often convenient to factor out certain dependences, yielding
\begin{align}
\Omega_{\rm GW}(\tau,k) 
= \kappa^2(\tau_*) \alpha_*^2 \left(\frac{\mathcal{H}_*}{\tilde{\beta}}\right)^2 
\lmk \frac{a^4 (\tau_*) \rho_{\rm tot}(\tau_*)}{a^4 (\tau) \rho_{\rm tot}(\tau)} \rmk 
\Delta(k, \tilde{\beta}) \,,
\end{align}
where $\kappa$ is the efficiency factor, $\alpha_*$ (assumed $\ll 1$) is defined by \eq{eq:alpha}, and
\begin{align}
\Delta(k, \tilde{\beta}) = \frac{3}{4\pi^2}\frac{\tilde{\beta}^2k^3}{\kappa^2(\tau_*) \rho_0^2(\tau_*)}
\int_0^{\tau_{\rm end}} d\tau_x
\int_0^{\tau_{\rm end}} d\tau_y \;
\lmk \frac{a (\tau_x) a(\tau_y)}{a^2(\tau_*)} \rmk^3 \cos(k(\tau_x - \tau_y))\Pi (\tau_x,\tau_y,k) \,.
\label{eq:DeltaApp}
\end{align}
In the main part of this paper, we take ${\tau_{\rm end}} \to \infty$ for notational simplicity. 
We note that the ratio $a^4 (\tau_*) \rho_{\rm tot}(\tau_*) / (a^4 (\tau) \rho_{\rm tot}(\tau))$ equals unity during the radiation-dominated epoch. However, we do not assume this throughout the paper unless explicitly stated.

The present-day GW spectrum can be obtained by accounting for the redshift after production.
Suppose that the GWs are generated during the radiation-dominated epoch.
The scale factor evolves according to
\begin{align}
\frac{a (\tau_*)}{a (\tau_0)}
&= 8.0\times10^{-16}
\left( \frac{g_{*s}}{100} \right)^{-\frac{1}{3}} \left(\frac{T_*}{100 \, \text{GeV}}\right)^{-1} \,,
\end{align}
where $T_*$ denotes the temperature at $\tau = \tau_*$ and $\tau_0$ is the present conformal time. The effective relativistic degrees of freedom for entropy and energy density are denoted by $g_{*s}$ and $g_*$. 
The physical frequency and GW amplitude at present are given by 
\begin{align}
f
&=
f_*\left(\frac{a (\tau_*)}{a (\tau_0)}\right) \nonumber \\
&\simeq 1.65 \times 10^{-5} \, {\rm Hz}
\left( \frac{f_*}{\beta} \right) \left( \frac{\beta}{H_*} \right)
\left( \frac{T_*}{10^2 \, {\rm GeV}} \right)
\left( \frac{g_*}{100} \right)^{\frac{1}{2}}
\left( \frac{g_{*s}}{100} \right)^{- \frac{1}{3}} \,,
\label{eq:f_present}
\end{align}
and
\begin{align}
\Omega_{\rm GW}h^2 &\simeq 1.67\times 10^{-5} \left( \frac{g_*}{100} \right) \left( \frac{g_{*s}}{100} \right)^{-\frac{4}{3}}
\Omega_{\rm GW}h^2\bigl|_{\tau = \tau_{\rm end}}\nonumber \\
&= 1.67\times 10^{-5}\kappa^2 (\tau_*) \alpha_*^2 \Delta \left( \frac{\beta}{H_*} \right)^{-2}
\left( \frac{g_*}{100} \right)
\left( \frac{g_{*s}}{100} \right)^{-\frac{4}{3}} \,.
\label{eq:Omega_present}
\end{align}
Here, we note that $f_* / \beta = k/(2 \pi \tilde{\beta})$ and $\beta / H_* = \tilde{\beta} / \mathcal{H}_*$.
If GWs are produced before reheating, additional dilution and redshift factors should be incorporated into the above expressions.

\bibliography{ref}

\end{document}